%FM 96-13, original archived in mp_arc 96-533, chao-dyn 96010???
%fiat.tex base
\newcount\mgnf\newcount\tipi\newcount\tipoformule
\newcount\aux\newcount\driver\newcount\pie\newcount\bibl
\newcount\xdata

\mgnf=0 
\tipoformule=0 
\aux=0   \def\9#1{\ifnum\aux=1#1\else\relax\fi}
\pie=1 
\bibl=1                 % 0= rif [XXX], 2= prepara, 1= rif. numerici
\xdata=1                %=0 giorno; =1 da scrivere

%%%%%%%%%%%%%%%%%%%%%%%%%%%%Incipit
\ifnum\mgnf=0
\magnification=\magstep0 
\hsize=13.5truecm\vsize=21truecm
\parindent=4.pt\fi
\ifnum\mgnf=1
\magnification=\magstep1
\hsize=16.0truecm\vsize=22.5truecm\baselineskip14pt\vglue6.3truecm
\parindent=4.pt\fi

%%%%%%%%%%%%%%%%%%%%%%%%%%%%Greco
\let\a=\alpha \let\b=\beta \let\g=\gamma \let\d=\delta 
\let\z=\zeta \let\h=\eta \let\th=\vartheta\let\k=\kappa \let\l=\lambda
\let\m=\mu \let\n=\nu \let\x=\xi \let\p=\pi \let\r=\rho
\let\s=\sigma \let\t=\tau  
 \let\o=\omega 
 \let\D=\Delta  \let\L=\Lambda
\let\P=\Pi    \let\O=\Omega

%%%%%%%%%%%%%%%%% Data
{\count255=\time\divide\count255 by 60 \xdef\oramin{\number\count255}
 \multiply\count255 by-60\advance\count255 by\time
 \xdef\oramin{\oramin:\ifnum\count255<10 0\fi\the\count255}}
\def\ora{\oramin }
\def\data{\number\day/\ifcase\month\or gennaio \or febbraio \or marzo \or
aprile \or maggio \or giugno \or luglio \or agosto \or settembre
\or ottobre \or novembre \or dicembre \fi/\number\year;\ \ora}
\setbox200\hbox{$\scriptscriptstyle \data $}
\ifnum\xdata=1\setbox200\hbox{$\scriptscriptstyle 27 ottobre 1996$}\fi
\newcount\pgn \pgn=1
\def\foglio{\number\numsec:\number\pgn
\global\advance\pgn by 1}
\def\foglioa{A\number\numsec:\number\pgn
\global\advance\pgn by 1}

%%%%%%%%%%%%%% Numerazione formule
\global\newcount\numsec\global\newcount\numfor
\global\newcount\numfig
\gdef\profonditastruttura{\dp\strutbox}
\def\senondefinito#1{\expandafter\ifx\csname#1\endcsname\relax}
\def\SIA #1,#2,#3 {\senondefinito{#1#2}
\expandafter\xdef\csname #1#2\endcsname{#3} \else
\write16{???? ma #1,#2 e' gia' stato definito !!!!} \fi}
\def\etichetta(#1){(\veroparagrafo.\veraformula)
\SIA e,#1,(\veroparagrafo.\veraformula)
 \global\advance\numfor by 1
\9{\write15{\string\FU (#1){\equ(#1)}}}
\9{ \write16{ EQ \equ(#1) == #1 }}}
\def \FU(#1)#2{\SIA fu,#1,#2 }
\def\etichettaa(#1){(A\veroparagrafo.\veraformula)
 \SIA e,#1,(A\veroparagrafo.\veraformula)
 \global\advance\numfor by 1
\9{\write15{\string\FU (#1){\equ(#1)}}}
\9{ \write16{ EQ \equ(#1) == #1 }}}
\def\getichetta(#1){Fig. \verafigura
 \SIA e,#1,{\verafigura}
 \global\advance\numfig by 1
\9{\write15{\string\FU (#1){\equ(#1)}}}
\9{ \write16{ Fig. \equ(#1) ha simbolo #1 }}}
\newdimen\gwidth

\def\BOZZA{
\def\alato(##1){
 {\vtop to \profonditastruttura{\baselineskip
 \profonditastruttura\vss
 \rlap{\kern-\hsize\kern-1.2truecm{$\scriptstyle##1$}}}}}
\def\galato(##1){ \gwidth=\hsize \divide\gwidth by 2
 {\vtop to \profonditastruttura{\baselineskip
 \profonditastruttura\vss
 \rlap{\kern-\gwidth\kern-1.2truecm{$\scriptstyle##1$}}}}}
}
\def\alato(#1){}
\def\galato(#1){}
\def\veroparagrafo{\number\numsec}\def\veraformula{\number\numfor}
\def\verafigura{\number\numfig}
\def\geq(#1){\getichetta(#1)\galato(#1)}
\def\Eq(#1){\eqno{\etichetta(#1)\alato(#1)}}
\def\eq(#1){\etichetta(#1)\alato(#1)}
\def\Eqa(#1){\eqno{\etichettaa(#1)\alato(#1)}}
\def\eqa(#1){\etichettaa(#1)\alato(#1)}
\def\eqv(#1){\senondefinito{fu#1}$\clubsuit$#1\write16{Manca #1 !}
\else\csname fu#1\endcsname\fi}
\def\equ(#1){\senondefinito{e#1}\eqv(#1)\else\csname e#1\endcsname\fi}
\openin13=#1.aux \ifeof13 \relax \else
\input #1.aux \closein13\fi
\openin14=\jobname.aux \ifeof14 \relax \else
\input \jobname.aux \closein14 \fi
\9{\openout15=\jobname.aux}

%%%%%%%%%%%%%%% Tipi
\newskip\ttglue%
\font\eighttt=cmtt8\font\sevenit=cmti7\font\sevensl=cmsl8%
\def\settepunti{\def\rm{\fam0\sevenrm}%
\textfont0=\sevenrm\scriptfont0=\fiverm\scriptscriptfont0=\fiverm%
\textfont1=\seveni\scriptfont1=\fivei\scriptscriptfont1=\fivei%
\textfont2=\sevensy\scriptfont2=\fivesy\scriptscriptfont2=\fivesy%
\textfont3=\tenex \scriptfont3=\tenex \scriptscriptfont3=\tenex%
\textfont\itfam=\sevenit\def\it{\fam\itfam\sevenit}%
\textfont\slfam=\sevensl\def\sl{\fam\slfam\sevensl}%
\textfont\ttfam=\eighttt\def\tt{\fam\ttfam\eighttt}%
\textfont\bffam=\sevenbf\scriptfont\bffam=\fivebf%
\scriptscriptfont\bffam=\fivebf \def\bf{\fam\bffam\sevenbf}%
\tt \ttglue=.5em plus.25em minus.15em%
\setbox\strutbox=\hbox{\vrule height6.5pt depth1.5pt width0pt}%
\normalbaselineskip=8pt\let\sc=\fiverm \normalbaselines\rm}%
\let\nota=\settepunti%
\font\titolo=cmbx10\font\titolone=cmbx12 scaled \magstep2%
\font\cs=cmcsc10\font\sss=cmss8%
\font\tenmib=cmmib10\font\sevenmib=cmmib10 scaled 800%

\textfont5=\tenmib  \scriptfont5=\sevenmib  %\scriptscriptfont5=\fivei

%%%%%%%%%%%%%% Grafica
\newdimen\xshift \newdimen\xwidth \newdimen\yshift  \newdimen\ywidth

\def\ins#1#2#3{\vbox to0pt{\kern-#2 \hbox{\kern#1 #3}\vss}\nointerlineskip}
\def\eqfig#1#2#3#4#5{
\par\xwidth=#1 \xshift=\hsize \advance\xshift 
by-\xwidth \divide\xshift by 2
\yshift=#2 \divide\yshift by 2
\line{\hglue\xshift \vbox to #2{\vfil
#3 \includegraphics{#4.ps}
}\hfill\raise\yshift\hbox{#5}}}
\def\8{\write13}

\def\didascalia#1{\vbox{\nota\0#1\hfill}\vskip0.3truecm}

\def\eqfigbis#1#2#3#4#5#6#7{
\par\xwidth=#1 \multiply\xwidth by 2
\xshift=\hsize \advance\xshift
by-\xwidth \divide\xshift by 3
\yshift=#2 \divide\yshift by 2
\ywidth=#2
\line{\hglue\xshift
\vbox to \ywidth{\vfil #3 \includegraphics{#4.ps}}
\hglue30pt
\vbox to \ywidth{\vfil #5 \includegraphics{#6.ps}}
\hfill\raise\yshift\hbox{#7}}}

\def\dimenfor#1#2{\par\xwidth=#1 \multiply\xwidth by 2
\xshift=\hsize \advance\xshift
by-\xwidth \divide\xshift by 3
\divide\xwidth by 2
\yshift=#2 %\divide\yshift by 2
\ywidth=#2}
 
\def\eqfigfor#1#2#3#4#5#6#7#8#9{
\line{\hglue\xshift
\vbox to \ywidth{\vfil #1 \includegraphics{#2.ps}}
\hglue30pt
\vbox to \ywidth{\vfil #3 \includegraphics{#4.ps}}\hfill}
%\vglue20pt
\line{\hfill\hbox{#9}}
\line{\hglue\xshift
\vbox to \ywidth{\vfil #5 \includegraphics{#6.ps}}
\hglue30pt
\vbox to\ywidth {\vfil #7 \includegraphics{#8.ps}}\hfill}}
  
\def\eqfigter#1#2#3#4#5#6#7{
\line{\hglue\xshift
\vbox to \ywidth{\vfil #1 \includegraphics{#2.ps}}
\hglue30pt
\vbox to \ywidth{\vfil #3 \includegraphics{#4.ps}}\hfill}
\multiply\xshift by 3 \advance\xshift by \xwidth \divide\xshift by 2
\line{\hfill\hbox{#7}}
\line{\hglue\xshift
\vbox to \ywidth{\vfil #5 \includegraphics{#6.ps}}\hfill}}

%%%%%%%%%%% Numer. bibliografia
%\newcount\bibl
%\bibl=0                % 0= rif [XXX], 2= prepara, 1= rif. numerici
\ifnum\mgnf=0\bibl=0\else\bibl=1\fi
\ifnum\bibl=0
\def\ref#1#2#3{[#1#2]}
\def\rif#1#2#3#4{\item{[#1#2]} #3}
\fi
\ifnum\bibl=2
\openout8=ref.b\openout9=fin.b
\def\ref#1#2#3{[#1#2]\write8{#1@#2}}\def\raf#1#2#3#4{}
\def\rif#1#2#3#4{\write9{\noexpand\raf{#1}{#2}{\noexpand#3}{#4}}}
\fi
\ifnum\bibl=1
\def\ref#1#2#3{[#3]}
\def\rif#1#2#3#4{}
\def\raf#1#2#3#4{\item{[#4]}#3}
\fi

%%%%%%%%%%%%% Locale

\def\V#1{{\,\underline#1\,}}
\def\T#1{#1\kern-4pt\lower9pt\hbox{$\widetilde{}$}\kern4pt{}}
\let\dpr=\partial\def\Dpr{{\V\dpr}}
\let\io=\infty\let\ig=\int
\def\fra#1#2{{#1\over#2}}\def\media#1{\langle{#1}\rangle}\let\0=\noindent
\def\guida{\leaders\hbox to 1em{\hss.\hss}\hfill}
\def\tende#1{\vtop{\ialign{##\crcr\rightarrowfill\crcr
\noalign{\kern-1pt\nointerlineskip}
\hglue3.pt${\scriptstyle #1}$\hglue3.pt\crcr}}}
\def\otto{\,{\kern-1.truept\leftarrow\kern-5.truept\to\kern-1.truept}\,}

\def\pagina{\vfill\eject}
\let\ciao=\bye

\def\st{\scriptscriptstyle}
\def\*{\vskip0.3truecm}
\def\lis#1{{\overline #1}}

\def\eg{\hbox{\it e.g.\ }}
\def\ap{\hbox{\it a priori\ }}
\def\ie{\hbox{\it i.e.\ }}
\let\==\equiv
\def\\{\hfill\break}
\def\fiat{{}}
\def\annota#1{\footnote{${}^#1$}}

\ifnum\aux=1\BOZZA\else\relax\fi
\ifnum\tipoformule=1\let\Eq=\eqno\def\eq{}\let\Eqa=\eqno\def\eqa{}
\def\equ{{}}\footline={\hss\tenrm\folio\hss}\fi
\ifnum\pie=0\footline={\hss\tenrm\folio\hss}\fi
\ifnum\pie=1\footline={\rlap{\hbox{\copy200}\ $\st[\number\pageno]$}
\hss\tenrm\folio\hss}\fi
\ifnum\pie=2\footline={\rlap{\hbox{\copy200}\ $\st[\number\pageno]$}
\hss\tenrm\foglio\hss}\fi

\fiat

\fiat
 
%\headline{\nota\hss To appear in ???}
\def\CC{{\cal C}}\def\AA{{\cal A}}\def\SC{{\cal S}}\def\EE{{\cal E}}
\def\W#1{#1_{\kern-3pt%
\lower7.5pt\hbox to 1truemm{\hfil$\widetilde{}$\hfil}}\kern2pt\,}
\def\DEF{\,{\buildrel def \over =}\,}
\def\EE{{\cal E}}\def\NN{{\cal N}}\def\KK{{\cal K}}
\def\1{\ifnum\mgnf=0\pagina\fi}
\def\KK{{\cal K}}
 
%%%%%%%%%%%%%%%%%%%%%%%%
 
\fiat
\vglue0.5truecm
\centerline{\titolone New methods in nonequilibrium}
\centerline{\titolone gases and fluids\annota{@}{\nota\rm
This paper is dedicated
to the memory of Giovanni Paladin, in grief for the untimely end of his
life, 29 june 1996.}}
\vskip1.truecm
\centerline{\titolo  Giovanni Gallavotti$^{*}$}
\vskip.5truecm
\centerline{\sss $^{*}$ Fisica, Universit\`a di Roma, P.le Moro 2,
00185 Roma}
\vskip1.truecm
 
\0{\cs Abstract.}
{\sss Kinematical and dynamical properties of chaotic systems are reviewed
and a few applications are described.}
 
\vskip1.truecm
 
\0{\it\S1 Kinematics.}
\numsec=1\numfor=1
\*
Gases and fluids are described mathematically by dynamical systems. This
means that the system state is a point in a {\it phase space} $\CC$ on
which the motion is given by a map $S$.
 
Typically $\CC$ is a finite dimensional smooth finite (\ie compact)
manifold and $S$ is a local diffeomorphism of $\CC$. But sometimes $\CC$
and/or $S$ have singularities (think for instance of hard sphere
systems) located on a closed set $N$ of zero volume: in such cases the
points of $\CC/N$ are the {\it regular points} where $S$ is smooth, of
class $C^\io$, with non vanishing jacobian determinant and $N$ will
always contain the boundary of $\CC$, if any.
 
The transformation $S$ can be {\it invertible}: this means that there is
another map $S^{-1}$ with a singularity set $N'$ (with $0$ volume) such
that if $x\not\in N$ and $Sx\not\in N'$ then $S^{-1}Sx=x$, and if
$x\not\in N'$ and $S^{-1}x\not\in N'$ then $S S^{-1}x=x$.
 
{\it Unless explicitly stated we always suppose the maps we consider to
be invertible and without singularities}.
 
If there is a smooth map $i$ of $\CC$ into itself such that $i^2=1$, and
$iS=S^{-1}i$ we say thgat the system is time reversal symmetric. We can
always suppose that $i$ is an isometry (possibly changing the distance
$d(x,y)$ to $\tilde d(x,y)=d(x,y)+d(ix,iy)$).
 
\*
 
The main question is: {\it given $(\CC,S)$ and picking $x\in\CC$
randomly with a distribution $\m_0$ proportional to the volume $dx$:
$\m_0(dx)=\r_0(x) \,dx$ (for some $\r_0\in L_1(\CC,dx)$), which is the
{\sl asymptotic behaviour} of the motion starting at $x$?}
 
This means asking what can one say about the averages:
 
$$\lim_{T\to\io}\fra1T\sum_{j=0}^{T-1} f(S^{\pm
j}x)=\media{f}_\pm\Eq(1.1)$$
In the case of a non invertible transformation one chooses, of course,
only the sign $+$ in \equ(1.1).
 
We shall always suppose that the averages exist for $\m_0$--almost all
$x$'s: this assumpiton is called {\it $0$-th law}. The necessity of the
clause "$\m_0$--almost everywhere" is because in essentially all cases
there are obvious exceptions.  Typically each periodic orbit is an
exception and usually periodic orbits form a dense set in phase space
(the cases with no periodic orbits, as the quasi periodic motions, must
be regarded as exceptional).
 
The independence on $x$ of the r.h.s. of \equ(1.1), \ie the uniqueness
of the averages, is not a very strong assumption. Usually it fails
because $\CC$ breaks into open parts $U_1,U_2,\ldots$ whose union has
full volume and in each of them the asymptotic behavior is unique, \ie
for each $j$ one has $x$--independence almost everywhere of the average
$\m_0$--almost everywhere in $U_j$. Thus the system would, essentially,
break down into several systems each of which verifies the $0$--th law.
 
The averages $\media{f}_\pm$ can be written:
 
$$\media{f}_\pm=\ig_\CC f(y)\m_\pm(dy)\Eq(1.2)$$
where $\m_\pm$ is a {\it probability distribution} that will be called
the  {\it forward statistics} ($+$) and the {\it backward statistics}
($-$) of $\m_0$ or the statistics of the motions with respect  to
initial data choices at random with respect to the volume in  phase
space.
 
Are there systems $(\CC,S)$ for which $\m$ can be determined?
\*
\0(1) {\it Hamiltonian systems}: in this case the phase space is the
space of the configurations and momenta $x$ of given energy that verify
some {\it "timing property"}, for instance $x$ lies on a certain surface
signaling some interesting event, \eg a collision.  If $t(x)$ is the
time interval between the "event" $x$ and the next one of the same type
the map $S$ is expresssed in terms of the solution flow $t\to S_t$ of
the equations of motion by $S=S_{t(x)}$.  The Liouville theorem gives
us a probability distribution $\m_0$ which is proportional to the
volume on $\CC$ and it is {\it invariant}: $\m_0(E)=\m_0(S^{-1}E)$ for
all sets $E$ for which it makes sense to measure the volume ("volume
measurable sets").
 
The {\it ergodic hypothesis}, when reasonable, tells us that $\m_0$ is
the statistics of itself: $\m_+=\m_0$, \ref{Ga}{1}{1}.
\*
 
\0(2){\it Transitive Anosov systems}: in this case the phase space is a
smooth manifold (say $C^\io$) and through each point $x$ pass a smooth
stable manifold $W^s_x$ and a smooth unstable manifold $W^u_x$ such that
their tangent planes $T^s_x, T^u_x$ at $x$:
\*
\0(i) are $S$--covariant (\ie $\dpr S\, T^u_x=T^u_{Sx},\dpr\,
S T^s_x=T^s_{Sx}$), vary continuously with $x$, in the sense that
$T^s_x,T^u_x$ are H\"older continuous as functions of
$x$,\annota{1}{\nota One might prefer to require real smoothness, \eg
$C^p$ with $1\le p\le \io$: but this would be too much for rather
trivial reasons.  On the other hand H\"older continuity might be
equivalent to simple $C^0$--continuity as in the case of Anosov systems,
see \ref{AA}{}{4}, \ref{Sm}{}{3}.} 
and the full tangent plane is the direct sum of
$T^s_x,T^u_x$. \\
\0(ii) for all $n\ge0$:
$$\eqalign{
(a)\ \ &||\dpr\,S^n \V v||_{S^nx}\le
C e^{-n\l}||\V v||_x\qquad \V v\in T^s_x\cr
(b)\ \ &||\dpr\,S^{-n} \V v||_{S^nx}\le C e^{-n\l}||\V v||_x\qquad \V v\in
T^u_x\cr}\Eq(1.3)$$
\0(iii) There is a periodic point with stable and unstable manifolds
dense in  $\CC$.
\*
%if $U,V$ are open then for all $n$ large enough $S^nU\cap V\ne\emptyset$.
 
It then follows that all points have dense stable and unstable manifolds
and that the periodic points are dense.
 
For such systems the $0$--th law holds as a mathematical theorem,
\ref{S}{}{2}, \ref{Sm}{}{3}, p. 757.
\*
 
\0(3) {\it Transitive Axiom A attractors}: Suppose that $(\CC,S)$ is a
smooth system and $\SC$ is a smooth invariant surface which is
attracting in the sense that the distance of $S^n x$ from $\SC$ tends
to $0$ exponentially: $d(S^n,\SC)\le C e^{-\l n}d(x,\SC)$ for all $x$
in the vicinity of $\SC$ (with $C,\l>0$ suitable constants).  If
$(\SC,S)$ is a transitive Anosov system then this is an example of a
{\it transitive Axiom A attractor}.  More generally an Axiom A
attractor needs not be a smooth manifold; it can be a non smooth closed
attracting set $\AA$ provided: \*
 
\0(a) at each point $x\in \AA$ the tangent space to $\CC$ can be
decomposed into two planes $T^s_x,T^u_x$ so that (i), (ii) above hold,
and (iii) is replaced by the requirement that on $\AA$ there is
a periodic point with dense stable and unstable manifolds.  \*
 
For such systems the $0$--th law holds and the statistics $\m_\pm$ are
called {\it SRB distributions}.
 
In general we call {\it attracting set} a closed invariant set $\AA$
such that $d(S^n x,\AA)\tende{n\to+\io}0$ for all $x$ sufficiently close
to $\AA$, and furthermore no subset of $\AA$ enjoys the same property.
A {\it repeller} is defined analogously by using $S^{-1}$ instead of
$S$.
 
Note that if $\AA$ is an Axiom A attractor then the planes $T^u_x$ are
"tangent to the attractor" in the sense that they are everywhere
tangent to a smooth invariant surface $W^u_x$ contained in $\AA$.  But
the plane $T^s_x$ are not tangent to $\AA$: in fact part of them sticks
out of $\AA$ (describing the fall on the attractor of the nearby
points) and part of them lies in some sense on the attractor and
describes motions that stay on the attractor and are {\it homoclinic}
as $n\to+\io$, \ie approach each other as $n\to+\io$.
 
In general through every point in an Axiom A attractor pass
two smooth manifolds $W^s_x,W^u_s$ which are tangent to $T^s_x,T^u_x$.
 
\*
 
\0(4) {\it Axiom A systems}: A notion more general than the previous
one can be given by introducing the {\it non wandering points}: they
are the points $x$ such that any of their neighborhoods returns
infinitely often near $x$.  This means that for any pair of
neighborhoods $U,V$ of $x$ there are infinitely many $n>0$ such that
$S^n V\cap U\ne\emptyset$, \ref{Sm}{}{3} p.749.
 
A smooth system $(\CC,S)$ verifies Axiom A if the set $\O$ of
nonwandering points is hyperbolic, \ie it verifies property (i), (ii)
of the previous example (2) and (iii) is replaced by the requirement
that the periodic points are dense, \ref{Sm}{}{3} p.777, \ref{R}{4}{5} p.
154.
 
If an Axiom A system has an attractive set then it is called
an ``axiom A attractor'' (not necessarily transitive, however see below).
In general through every non wandering point in an Axiom A system pass
two smooth manifolds $W^s_x,W^u_s$ which are tangent to $T^s_x,T^u_x$.
 
The nonwandering set of an Axiom A system splits into a finite union of
invariant sets, called {\it basic sets}, $\{\O_j\}$ on each of which
there is a dense orbit ("topological transitivity", not to be confused
with the above notion of transitivity). And each of such sets splits
into a finite union of sets $\{B_{jk}\}$ each of which is invariant for
an appropriate {\it iterate} $S^{p_j}$ which acts transitively on it
(in the sense of the previous examples: namely the stable and unstable
manifolds of a periodic point are dense). Thus if $\O_j$ is an attractor
each $B_{jk}$ is an Axiom A transitive attractor for $S^{p_j}$).
 
The $0$--th law holds for the attractors of an Axiom A system.
\*
\0(5) {\it Axiom C systems}: The Axiom A and the zeroth law properties
could be reasonably taken as properties characterizing models for
globally "chaotic" or "globally hyperbolic" systems.  But in Axiom A
systems there are no relations between the different basic sets.  And
therefore other more global kinematic notions have been considered in
the literature: we shall not dwell here on the Axiom B notion and we
deal directly with the Axiom C notion that is particularly appropriate
for invertible maps enjoying a time reversal symmetry.  In fact the
problems that we pose in the next sections suggest that the appropriate
notion for "globally hyperbolic" or "globally chaotic" dynamical
systems is somewhat stronger than that of Axiom A or B.
 
Axiom C systems are Axiom A systems enjoying further properties.  To
describe them we introduce the notion of distance of a point $x$ to the
basic sets $\{\O_i\}$ of an Axiom A system as:
$$\d(x)=\min\{\min_i \fra{d_{\O_i}(x)}{d_0}, \inf_{j,\,-\io<n<+\io}
\fra{d_{\O_j}(S^nx)}{d_0}\}\Eq(1.4)$$
where $d_0$ is the diameter of the phase space $\CC$, $d_{\O_i}(x)$ is
the distance of the point $x$ from the basic set $\O_i$, the
minimum over $i$ runs over the attracting or repelling basic sets and
the minimum over $j$ runs over the other basic sets (if any).
 
We can then define the {\it Axiom C} systems as:
\*
 
\0A smooth dynamical system $(\CC,S)$ verifies Axiom C if it verifies
Axiom A and:
 
\0(1) among the basic sets there are a unique attracting and a unique
repelling basic sets, denoted $\O_+,\O_-$ respectively, with (open)
full volume dense basins of attraction. We call $\O_\pm$ the {\it poles}
of the system (future or attracting and past or repelling poles,
respectively).
 
\0(2) for every $x\in\CC$ the tangent space $T_x$ admits a
H\"older--continuous decomposition as a direct sum of three subspaces
$T^u_x,T^s_x,T^m_x$ such that:
 
\halign{\qquad
#\quad&#           &#     \hfill              &\qquad#\hfill\cr
a)& $\dpr S\, T^\alpha_x$&$=T^\alpha_{Sx}$          &$\a=u,s,m$ \cr
b)& $|\dpr S^n w|$     &$\le C e^{-\l n}|w|$,     &$w\in T^s_x,\ n\ge0$\cr
c)& $|\dpr S^{-n} w|$  &$\le C e^{-\l n}|w|$,     &$w\in T^u_x,\ n\ge0$\cr
d)& $|\dpr S^n w|$     &$\le C \d(x)^{-1} e^{-\l |n|}|w|$,&$w\in T^m_x,\
\forall n$\cr}
 
\0where the dimensions of $T^u_x,T^s_x, T^m_x$ are $>0$ and $\d(x)$ is
defined in \equ(1.4); here $\dpr S^n$ is the jacobian matrix of $S^n$.
 
\0(3) if $x$ is on the attracting pole $\O_+$ then $T^s_x\oplus
T^m_x$ is tangent to the stable manifold in $x$; viceversa if $x$ is on
the repelling pole $\O_-$ then $T^u_x\oplus T^m_x$ is tangent to the
unstable manifold in $x$. \*
 
Although $T^u_x$ and $T^s_x$ are not uniquely determined by the above
definition the planes $T^s_x\oplus T^m_x$ and $T^u_x\oplus T^m_x$ are
uniquely determined for $x\in\O_+$ and, respectively, $x\in\O_-$.
 
The above notion of Axiom C is in \ref{BG}{}{6}.
It is clear that an Axiom C system is necessarily also an Axiom A
system verifying the $0$--th law.
 
If $\O_+$ and $\O_-$ are the two poles of the system the stable
manifold of a periodic point $p\in \O_+$ and the unstable manifold of a
periodic point $q\in\O_-$ not only have a point of transversal
intersection (this would be the property characterizing Axiom B
systems among the Axiom A ones with only two basic sets) but they
intersect transversally {\it all the way} on a manifold connecting
$\O_+$ to $\O_-$; the unstable manifold of a point in $\O_-$ will
accumulate on $\O_+$ {\it without winding around it}, \ref{BG}{}{6}.
 
In fact one can {\it "attach"} to $W^s_p$, $p\in\O_+$, points on $\O_-$
as follows: we say that a point $z\in \O_-$ is {\it attached} to $W^s_p$
if it is an accumulation point for $W^s_p$ and there is a closed curve
{\it with finite length} linking a point $z_0\in W^s_p$ to $z$ and {\it
entirely lying on $W^s_p$}, with the exeption of the endpoint $z$.  A
drawing helps understanding this simple geometrical construction,
slightly unusual because of the density of $W^s_p$ on $\O_-$,.
 
We call $\lis W^s_p$ the set of the points {\it either on $W^s_p$ or
just attached to $W^s_p$} on the system basic sets (the set $\lis
W^s_p$ should not be confused with the closure ${\rm clos}(W^s_p)$,
which is the whole space, see \ref{Sm}{}{3}, p.  783).  If a system
verifies Axiom C the set $\lis W^s_p$ intersects $\O_-$ on a stable
manifold, by 2) in the above definition.
 
The definition of $\lis W^u_q$, $q\in \O_-$, is set symmetrically by
exchanging $\O_+$ with $\O_-$.
 
Furthermore if a system verifies Axiom C and $p\in \Omega_+$, $q\in
\O_-$ are two periodic points, on the attracting and on
the repelling pole of the system respectively, then $\lis W^s_p$
and $\lis W^u_q$ have a dense set of points in $\Omega_-$ and
$\Omega_+$, respectively.
 
Note that $\lis W^s_p\cap\lis W^u_q$ is dense in $\CC$ as well as in
$\Omega_+$ and $\Omega_-$.  This follows from the density of $W^s_p$
and $W^u_q$ on $\Omega_+$ and $\Omega_-$ respectively and from the
continuity of $T^m_x$.  Furthermore if $z\in \Omega_+$ is such that
$z\in \lis W^s_p\cap\lis W^u_q $ then the surface $\lis
W^s_p\cap\lis W^u_q$ intersects $\Omega_-$ in a unique point $\tilde
z\=\tilde\imath z$ which can be reached by the shortest smooth path on
$\lis W^s_p\cap\lis W^u_q$ linking $z$ to $C_-$ (the path is on the
surface obtained as the envelope of the tangent planes $T^m$, but it is
in general not unique even if $T_m$ has dimension $1$, see the example
in Appendix A1 below).
 
The map $\tilde\imath$ commutes with $S$, squares to the
identity, maps $\O_+$ to $\O_-$ and viceversa
and will play a key role in the following analysis.
 
In \ref{BG}{}{6} it is conjectured that the Axiom C systems are
$\O$--stable in the sense of Smale, \ref{Sm}{}{3} p.  749, \ie small
perturbations of Axiom C systems are still Axiom C systems.\*
 
\0{\it\S2 Dynamics and chaotic hypothesis}
\numsec=2\numfor=1\*
 
We begin with a few examples of systems that we would like to study.
 
\0{\it Example 1}:
We shall consider the problem of electrical conduction in a crystal
via the classical model  representing the crystal as a periodic array
of circular obstacles among which free charges, also modeled by hard
spheres, move undergoing elastic collisions with the obstacles and
between themselves.
 
An electric field acts upon the moving particles establishing a current.
We suppose that the obstacles are such that there is no straight line
that avoids them. Nevertheless clearly the electric field will continue
to work so that the electric current will grow unbounded and therefore
the system will show infinite conductivity.
 
The reason why electrical conductivity is finite in physical systems is
simply that there are {\it dissipative effects}. The simplest theories
of conductivity, like Drude's theory, write directly a dissipative
equation for the motion of the $N$ moving particles contained in a
crystal with cell of size $a$:
 
$$
\dot{\V q}_j
=\fra{\V p_j}m,\qquad
\dot{\V p}_j=\V F_j+E\V i-\n \V p_j\Eq(2.1)$$
where $\n=2\ell v^{-1}$ if $\ell$ is the mean free path (equal to
$\ell=(4\p\r d^2)^{-1}$ if $d$ is the radius of the particles and
$\r=N/a^3$ their density) and $v$ is the average velocity, $F_j$ is the
(impulsive) force acting on the $j$--th particle. This special choice of
$\n$ corresponds to a dissipation proportional to the speed: no special
physical meaning should be attached to this choice which is here used
only as an example.
 
Periodic boundary conditions are imposed at the cell boundary in the
direction parallel to the field and reflecting conditions are imposed
in the direction orthogonal to the field ({\it semiperiodic boundary
conditions}), to fix the ideas.
 
The above are not the only equations that we can use for the conduction
problem: in fact the coefficient $\n$ has an empirical nature and it is
supposed to model a thermostat action that forbids the system to reach
arbitrarily high current.
 
At the same level of rigor one could equally well maintain that the
thermostat action is that of keeping the total energy of the moving
particles constant; hence one can think of imposing the anholonomic
constraint $\sum_j \V p_j^2=const$ and the constant should be
$\EE=N(\fra32 k_BT+\fra12 m\lis v^2)$ if $T$ is the absolute
temperature, $k_B$ the Boltzmann constant and $\lis v$ is the average
drift velocity.
 
Gauss asked himself what would be the way of imposing a nonholonomous
constraint affecting in a "minimal fashion" the dynamics and formulated
the well known extension of D'Alembert's principle called the {\it
least constraint} principle.  If the constraint of constant energy is
imposed on the above system, by Gauss' principle, the resulting
equations take the form \equ(2.1) with $\n$ replaced by a multiplier
$\a$ given by:
 
$$\a(\V p)=\fra{E\V i\cdot\sum_j\V p_j}{\sum_j\V p_j^2}\Eq(2.2)$$
A brief computations shows that indeed if in \equ(2.1) one substitutes
$\n\to\a$ the solutions of the new equation \equ(2.1) keep a constant
total (kinetic) energy. This model was considered for instance in
\ref{GC}{2}{7}.
 
Of course the value of the constant given to the energy cannot be
arbitrary.  Suppose that the dissipative effects modeled by $\n$ are
such that when the external field is $E$ then the system relaxes to a
stationary state in which the average energy is $\EE$.  Then if we want
that the model \equ(2.1) or the same model with $\n$ replaced by the
$\a$ of \equ(2.2) to be equivalent we should fix the energy in the
second model to be exactly $\EE$.  For large systems the two models
should be equivalent, \ref{Ga}{6}{8}.
\*
 
\0{\it Example 2}:
A more physical model for electric conduction is:
 
$$\dot{\V q}_j=\V p_j/m,\qquad
\dot{\V p}_j=\V F_j-\b\V p_j+m\dot{\V B}_j(t)+ E\V i\Eq(2.3)
$$
where $\V B_j$ is a brownian motion. This is a {\it Langevin equation}
in which $\b$ is a friction coefficient (\eg $\n$ above) and the
dispersion $\d$ of each component of $\dot{\V B}_j$ is related to the
temperature (by $\fra12\fra{m\d^2}{2\b}=\fra12 k_B T$, see \ref{N}{}{9},
p. 55).
 
The above equation is clearly phenomenological and one should think that
the same physical system may be described in a different way, for
instance by the system of equations:
 
$$\eqalign{
&\cases{\dot{\V q}_j=\V p_j/m\cr
\dot{\V p}_j=\V F_j-\a\V p_j+m\dot{\V B}_j(t)+ E\V i\cr},
\qquad
\cases{
\dot{\V b}_{jk}=\V B_{jk}\cr
\dot{\V B}_{jk}=-\Dpr_{\V b_{jk}} V(\{\V b_{j'k'}\})\cr}\cr
&\V B_j\DEF \sum_k \V B_{jk}\cr}\Eq(2.4)$$
where the multiplier $\a$ is determined by imposing that the total
kinetic energy $\EE=\sum_j \fra1{2m}\V p_j^2$
of the charged particles stays constant:
$$\a=\fra{\sum_j(E\V i-\dot{\V B}_j)\cdot\V p_j}{\sum_j\V
p_j^2}\Eq(2.5)$$
and the $(\V b,\V B)$ coordinates describe a hamiltonian system with
many degrees of freedom (\ie the label $k$ takes many values, \eg $k\gg
N$) which acts as a {\it heat reservoir} on the system of particles.
 
Again the equivalence between the two equations can hold only if the
value $\EE$ of the energy in \equ(2.4) is fixed equal to the average
value of the energy of \equ(2.3) and, at the same time, the motion of the
(hamiltonian) dynamical system for the evolution of the variables $\V
b,\V B$ is so chaotic that the variables $\V B_j$ can be reagrded as a
gaussian process whose covariance is the same, or close, to  that of the
homomimous stochastic process in \equ(2.3), \ref{Ga}{6}{8}.
\*
 
\0{\it Example 3} As third and last example we   shall focus on fluid
mechanics problems considering a fluid that:
 
\0(1) is enclosed in a periodic box $\O$ with side $L$, possibly
with a few disks ("obstacles") removed so that no infinite straight
path can be found in $\O$ that avoids the obstacles,
 
\0(2) is incompressible with density $\r$.
 
I shall consider two distinct evolution equations for this fluid, both
of dissipative nature.
 
$$\eqalign{
\dot{\V u}+\W u\cdot\W \dpr \,\V u=-\fra1\r\V\dpr p+\V g+\n\D\V u,
\qquad\V\dpr\cdot\V u=0&\qquad {\rm NS}\cr
\dot{\V u}+\W u\cdot\W \dpr \,\V u=-\fra1\r\V\dpr p+\V g+\b\D\V u,
\qquad\V\dpr\cdot\V u=0&\qquad {\rm GNS}\cr
}\Eq(2.6)$$
 
{\it In the case $\O$ contains obstacles a "no friction" boundary
condition will be imposed on $\dpr\O$,} \ie $\V u\cdot\V n=0$ if $\V n$
is the normal to $\dpr\O$.  The first equation is the well known Navier
Stokes equation with $\n$ being the {\it viscosity}.
 
The second equation, introduced in \ref{Ga}{6}{8} and called the
gaussian Navier Stokes equation or GNS equation, has a
multiplier $\b$ defined so that the total vorticity $\h L^3=\r\ig
\V\o^2\,dx$, with $\V\o=\Dpr\wedge\V u$ being the {\it vorticity}, is a
constant of motion; this means that:
 
$$\b(\V u)=\fra{\ig\big(\V\dpr\wedge
\V g\cdot\V\o+ \V\o\cdot \,(\W \o\cdot\W \dpr\V
u)\big)\,d\V x}{\ig(\Dpr\wedge\V\o)^2\,d\V x}\DEF \b_e+\b_i\Eq(2.7)$$
The above equations were studied in \ref{Ga}{7}{10}.
We shall consider only the {\it truncated equations} with momentum cut
off $K$. In this way the existence and uniquenes problems are completely
avoided.
 
The truncation is performed on a suitable orthonormal basis for
the {\it divergenceless} fields in $\O$: given the boundary conditions
we consider it natural to use the basis generated by the {\it minimax}
principle applied to the Dirichlet quadratic form $\ig_\O (\W\dpr\,\V
u)^2d\V x$ defined on the space of the $C^\io(\O)$ divergenceless
fields $\V u$ with $\V u\cdot\V n=0$ on $\dpr\O$.  The basis fields $\V
u_j$ will verify: $\D\V u_j=-E_j\V u_j+\V\dpr_j\m$, for a suitable
multiplier $\m_j$, with $\V u_j,\m_j\in C^\io$ and $E_j$ are
eigenvalues). Thus truncating at momentum $K$ means setting identically
$0$ the components on $\V u_j$ with $E_j> K^2$.
 
For instance in the case of {\it no obstacles} let the $\V u=\sum_{\V
k\ne\V0}\V\g_{\V k} e^{i\V k\cdot\V x}$ be the velocity field
represented in Fourier series with $\V \g_{\V k}=\lis{\V \g_{-\V k}}$
(reality condition) and $\V k\cdot\V\g_{\V k}=0$ (incompressibility
condition); here $\V k$ has components that are integer multiples of the
"lowest momentum" $k_0=\fra{2\p}L$.  Then consider the equation:
 
$$\dot{\V\g}_{\V k}=-\th(\V k)\V \g_{\V k} -i\sum_{\V k_1+\V k_2=\V k}
(\V\g_{\V k_1}\cdot\V k_2)\, \P_{\V k}\V \g_{\V k_2}+\V g_{\V k}
\Eq(2.8)$$
where the $\V k$'s take only the values $0<|\V k|<K$ for some {\it
momentum cut--off} $K>0$ and $\P_{\V k}$ is the projection on
the plane orthogonal to $\V k$. This is an equation that defines a
"truncation on the momentum sphere with radius $K$ of the equations
\equ(1.1)" if:
 
$$\eqalign{
\th(\V k)=-\n \V k^2\qquad & {\rm  NS\ case}\cr
\th(\V k)=-\b \V k^2\qquad & {\rm GNS\ case}\cr
}\Eq(2.9)$$
For simplicity we may suppose, in this no obstacles cases, that the mode
$\V k=\V 0$ is {\it absent}, \ie $\V\g_{\V0}=\V0$: this can be done if,
as we suppose, the external force $\V g$ does not have a zero mode
component (\ie if it has zero average).
 
In order that the resulting cut--off equations be physically acceptable,
and supposing that $\V g_\V k\ne\V0$ only for $|\V k|\sim k_0$, one shall
have to fix $K$ large. For instance in the NS case it should be
much larger than the {\it Kolmogorov scale} $K=(\h\n^{-2})^{1/4}$,
where $\n\h$ is the average dissipation of the solutions to \equ(2.6)
with $K=+\io$ (determined on the basis of heuristic dimensional
considerations by $\h\sim {|\V g|^2L^2\n^{-2}}$): see \ref{LL}{}{11}.
 
It is easy, in the no obstacles cases, to express the coefficients
$\b$ for the cut off equations as $\b\=\b_e+\b_i$:
 
$$\eqalign{
&\b_e=\fra{\sum_{\V k\ne\V0}\V k^2\V g_{\V k}\cdot
\lis{\V \g}_{\V k}}{\sum_{\V k}\V k^4|\V \g_{\V k}|^2}\cr
&\b_i=\fra{-i\sum_{\V k_1+\V k_2+\V k_3=\V0}
\V k_3^2\,(\V\g_{\V k_1}\cdot\V k_2)\,(\V \g_{\V k_2}\cdot\V \g_{\V
k_3})}{\sum_{\V k}\V k^4|\V \g_{\V k}|^2}\cr}\Eq(2.10)$$
where the $\V k$'s take only the values $0<|\V k|<K$ for some {\it
momentum cut--off} $K>0$ and $\P_{\V k}$ is the orthogonal projection on
the plane perpendicular to $\V k$.
 
The above two equations should be again {\it equivalent} in the natural
sense that the stationary distribution $\m_{\n,ns}$ of the (truncated)
NS equation with viscosity parameter $\n$ and the stationary
distribution $\m_{\h,gns}$ of the GNS equations should give the same
averages to "most observables" if the value $\h$ is fixed to be the same
as the average of the dissipation in NS equations, \ref{Ga}{7}{10}.
 
\*
 
{\it A key remark suggested by the above equivalence statements, which
are in fact {\it only conjectures}, is that the stationary distributions
can be regarded as statistical ensembles and the conjectures state that
the same thermostatting mechanism can be equivalently modeled by
completely different equations.}
 
Furthermore the same thermostatting mechanism can be modeled by a
"usual" irreversible equation of motion and {\it also} by a reversible
one.
 
Much in the same way as in equilibrium statistical mechanics the same
system can be described by the microcanononical or the canonical
ensembles. Of course as in classical equilibrium mechanics the ensembles
are not completely equivalent as one can always find quantities that
have a completely different behavior in two "equivalent" systems (\eg
the energy fluctuations are quite different in the canonical and
microcanonical ensembles even when they are equivalent).
 
The real problem is, however, {\it how to find the stationary
distributions}?
 
This has been an outstanding problem in the last century: a solution for
it was proposed in 1973 by Ruelle but it went unnoticed. Probably
because it was formulated (later, \eg in \ref{R}{2}{12}, and) in a way
that a concrete application seemed far away.
 
Recently the Ruelle's principle has been shown to imply far reaching
consequences. We therefore proceed to formulate it in a modern form
having in mind to apply it (as an example) to the systems
$(\CC,S)$ obtained by timed observations on the above described family
of systems.
 
In \ref{GC}{1}{13},\ref{GC}{2}{7},\ref{BGG}{}{14} the above systems, at
least when {\it chaotic i.e.} when showing at least one positive
Lyapunov exponent, are supposed to verify following hypothesis: \*
 
{\it Chaotic hypothesis: A reversible many particle system in a
stationary state can be regarded as a transitive Anosov system for the
purpose of computing the macroscopic properties.}
\*
 
This means that the attractor verifies Axiom A and it is assumed to be
smooth and hyperbolic in the strict mathematical sense described in the
kinematics section.  At zero forcing we suppose, for compatibility with
the ergodic hypothesis, the system to be a transitive hamiltonian Anosov
system so that the attractor is the full phase space.
 
In the quoted references it is argued that the chaotic hypothesis should
be considered in the same way as the {\it ergodic hypothesis} in
equilibrium statistical mechanics.  It is assumed as correct {\it even}
in cases in which it cannot be mathematically strictly valid: but this
can be done only for the purpose of deriving statistical properties of a
few relevant observables.  An analogue of this procedure is the
derivation of the second law ({\it heat theorem}) from ergodicity ({\it
i.e.} from the microcanonical ensemble) in equilibrium statistical
mechanics: the law is derived by supposing ergodicity and it is assumed
valid even when the ergodic hypothesis is obviously false ({\it e.g.}
for the free gas in a box), \ref{Ga}{1}{1}.
 
It might be surprising that something can be concretely derived from
this assumption: in fact it is very ambitious and far reaching and we
shall see it has implications.
 
And it is clear that any consequence of such a general assumption must
be a parameterless prediction, \ie a {\it universal law}. Examples of
this type of deductions are well known: the most remarkable is perhaps
the mentioned {\it heat theorem} of Boltzmann which derives the second
law of thermodynamics from the ergodic hypothesis (\ie from the
microcanonical ensemble).
 
In nonequilibrium thermodynamics there are no universally accepted laws
that we could use as a reference and a test of the theory: except perhaps
the Onsager reciprocity and the fluctuation dissipation theorems. The
latter are derived in very many different ways and there is a general
agreement about their correctness, see \ref{DGM}{}{15}.
 
The Onsager reciprocal relations and the fluctuation dissipation
relations (also called Green-Kubo formulae) have been subject of many
unsuccessful, or only partially successful, attempts to extend them to
non zero (or large) external forcing. In fact they "only" express
relations between derivatives with respect to external fields {\it
evaluated at $0$ fields}!
 
We shall argue that the above chaotic hypothesis not only implies the
Onsager relations (when they are valid, \ie at zero fields) but that it
also leads to an extension of them, in the form of a general
parameterless statement, that will be called the {\it fluctuation
theorem}.
 
\*
 
\0{\it\S3 Reversible dissipation and the fluctuation theorem. Markov
partitions (coarse graining and symbolic dynamics). }
\numsec=3\numfor=1\*
 
Note that the examples in \S2 are pairs of dynamical systems which are
{\it conjectured} to represent the statistical properties of the same
system and the pairs have been deliberately selected so that each pair
consists of a manifestly dissipative equation and of a manifestly
reversible equation.
 
It is easy to check that the second equation of the example 1) is
reversible: the operation $i:(\V p,\V q)\to(-\V p,\V q)$ induces a time
reversal in phase space in the sense that the solution flow $S_t$
verifies $iS_t=S_{-t}i$. This implies that the timing map $S$ (on any
time independent timing event) verifies $iS=S^{-1}i$.
 
Likewise the map $i:(\V p,\V B,\V q,\V b)\to (-\V p,-\V B,\V q,\V b)$
induces a time reversal symmetry in the phase space of the second
equation in the second example.
 
And the second equation of the third example admits also the time
reversal symmetry $i:\V u\to-\V u$.
 
In this section we shall restrict attention to such reversible systems
and we shall suppose that $i$ is an isometry (not restrictive, see \S1)
and that the chaotic hypothesis is verified.
 
Let $x$ be a point on the phase space $\CC$ of the systems that we study
through timed observations and therefore are dynamical systems $(\CC,S)$
of the type considered in \S1. We define the jacobian matrix $J(x)=\dpr
S(x)$ and regard it as a linear map of the tangent plane $T_x$ onto
$T_{Sx}$.  We also define the action of $\dpr S(x)$ on the stable and
unstable planes $T^u_x,T^s_x$ as a linear map $J_u(x)$ or $J_s(x)$ onto
$T^u_{Sx},T^s_{Sx}$.
 
We call $\L_u(x),\L_s(x)$ the determinants of the jacobians $J_u(x),
J_s(x)$: their product differs from the determinant $\L(x)$ of $\dpr
S(x)$ by the ratio of the sine of the angle $a(x)$ between the planes
$T^s_x,T^u_x$ and the sine of the angle $a(Sx)$ between
$T^s_{Sx},T^u_{Sx}$.  We set:
 
$$\L_{u,\t}(x)=\prod_{j=-\t/2}^{\t/2-1}\L_u(S^jx),\quad
\L_{s,\t}(x)=\prod_{j=-\t/2}^{\t/2-1}\L_s(S^jx),\quad
\L_{\t}(x)=\prod_{j=-\t/2}^{\t/2-1}\L(S^jx)\Eq(3.1)$$
hence $\L(x)=\fra{\sin a(Sx)}{\sin a(x)}\L_s(x)\L_u(x)$.
 
The time reversal symmetry implies that $W^s_{ix}=iW^u_x,
W^u_{ix}=iW^s_x$ and:
 
$$\eqalign{
&\L_\t(x)=\L_\t(ix)^{-1},\qquad \L_{s,\t}(ix)=\L_{u,\t}(x)^{-1},
\qquad \L_{u,\t}(ix)=\L_{s,\t}(x)^{-1}\cr
&\sin a(x)=\sin a(ix)\cr}\Eq(3.2)$$
 
The quantity $\s(x)=-\log \L(x)$ will be called the {\it entropy
production} per timing event so that $e^{-\s(x)}$ is the {\it phase space
volume contraction} per event.
 
{\it It seems that $\s(x)$ has all the properties that one may wish for
the extension to non equilibrium thermodynamics of the ordinary
entropy}, even though not everybody would agree on this statement.
 
For instance it is well established in many numerical experiments that
the time average of $\s(x)$ is $\ge0$ and $>0$ when the external fields
are non zero. In zero field $\s(x)$ {\it vanishes} (an expression of
Liouville's theorem).
 
Recently the general non negativity of the time average of $\s(x)$ has
become a theorem (by Ruelle) that can be rightly called, I feel, the
{\it H--theorem} of reversible non equilibrium statistical mechanics,
\ref{R}{3}{16}.  More generally one can consider systems that are smooth
only piecewise and extend the notion of SRB distribution so that, if the
latter does not have nontrivial vanishing Lyapunov exponents,
it can be also shown, {\it c.f.r} \ref{R}{3}{16}, that the average of
$\s(x)$ can vanish on a SRB distribution\annota{2}{\nota In
general there may be many of them but, by the chaotic hypothesis
assumed here, there is one and only one.} has a density with respect to
the volume.
 
Therefore we shall  call {\it dissipative} the systems for which the
time average of $\s(x)$  is positive, \ref{Ga}{2}{17}.
 
We call $\s_\t(x)$ the partial average of $\s(x)$ over the part of
trajectory centered at $x$ (in time): $S^{-\t/2}x,\ldots,
S^{\t/2-1}x$. Then we can define the {\it dimensionless entropy
production} $p=p(x)$ via:
 
$$\s_\t(x)=
\fra1\t\sum_{j=-\t/2}^{\t/2-1}\s(S^j x)\DEF \media{\s}_+ p\Eq(3.3)$$
where $\media{\s}_+$ is the infinite time average $\ig_\CC
\s(y)\m_+(dy)$, if $\m_+$ is the forward statistics of the volume
measure, and $\t$ is any integer.
 
The {\it chaotic hypothesis} implies a {\it fluctuation theorem} which,
\ref{GC}{2}{7}, is a property of the fluctuations of the entropy
production rate in dissipative systems; the (dimensionless) finite time
average $p=p(x)$ has a statistical distribution $\pi_\tau(p)$ with
respect to the stationary state distribution $\m_+$ such that the
following limit exists:
 
$$\lim_{\t\to\io}\fra1{\t\media{\s}_+}\log \p_\t(p)=-\z(p)\Eq(3.4)$$
for all $p$'s in the domain $(-p^*,p^*)$ where $p(x)$ can vary, and:
$$\fra{\z(p)-\z(-p)}{p\media{\s}_+}=-1\Eq(3.5)$$
provided the attractor is dense on phase space {\it and} the system is
reversible. This is the {\it fluctuation theorem} of \ref{GC}{1}{13}:
later we indicate its extension to cases in which the attractor is
smaller than the whole phase space. Under our assumptions the function
$\z(p)$ exists and is real analytic in $p\in(-p^*,p^*)$, \ref{Ga}{2}{17}.
 
The above relation holds in the whole domain of variability of $p$
(which is in general a bounded variable because $\CC$ is a bounded
manifold): and therefore it is a surprising parameterless prediction.
 
The proof of the fluctuation theorem goes to the core of the structure
of chaotic systems and is very enlightening. The key point is that
Anosov (and Axiom A) systems have a dynamics that can be {\it easily}
transformed into a symbolic dynamics, \ref{S}{}{2}.
 
One can in fact find a partition of phase space $\CC$ into {\it
parallelograms}. A parallelogram will be a small set with a boundary
consisting of pieces of the stable and unstable manifolds joined
together as described below. The smallness has to be such that the parts
of the manifolds involved look essentially ``straight'': \ie the sizes
of the sides have to be small compared to the smallest radii of
curvature of the manifolds $W^u_x$ and $W^s_x$, as $x$ varies in $\CC$.
 
Therefore let $\d$ be a length scale small compared to the minimal
(among all $x$) curvature radii of the stable and unstable manifolds.
Let $W^{u,\d}_x,W^{s,\d}_x$ be the connected parts of $W_x^u$, $W^s_x$
containing $x$ and contained in a sphere of radius $\d$.
Let us first define a {\it parallelogram} $E$ in the phase space $\CC$,
to be denoted by $\D^u\times\D^s$, with center $x$ and axes $\D^u$,
$\D^s$ with $\D^u$ and $\D^s$ small connected surface elements on
$W^u_x$ and $W^s_x$ containing $x$. Then $E$ is defined as
follows.
 
Consider $\x\in\D^u$ and $\h\in\D^s$ and suppose that the
intersection $\x\times\h\=W^{s,\d}_\x\cap W^{u,\d}_\h$ is a unique point
(this will be so if $\d$ is small enough and if $\D^u$, $\D^s$ are small
enough compared to $\d$ as we can assume, because the stable and
unstable manifolds are ``smooth''
and transversal).
 
The set $E=\D^u\times\D^s$ of all the points generated in this way when
$\x,\h$ vary arbitrarily in $\D^u,\D^s$ will be called a parallelogram
(or rectangle), if the boundaries $\dpr\D^u,\dpr\D^s$ of $\D^u$ and
$\D^s$ as subsets of $W^u_x$ and $W^s_x$, respectively, have zero
surface area on the manifolds on which they lie.  The sets $\dpr_u
E\=\D^u\times\dpr\D^s$ and $\dpr_s E=\dpr\D^u\times\D^s$ will be called
the {\it unstable} or {\it horizontal} and {\it stable} or {\it
vertical} sides of the parallelogram $E$.
\*
 
\eqfig{260pt}{90pt}{
\ins{43pt}{37pt}{$x$}
\ins{43pt}{60pt}{$\D^s$}
\ins{60pt}{40pt}{$\D^u$}
\ins{155pt}{36pt}{$\x$}
\ins{130pt}{60pt}{$\h$}
\ins{177pt}{77pt}{$\x\times\h$}
\ins{245pt}{70pt}{$E$}
}{c3gcpaper}{}
\*
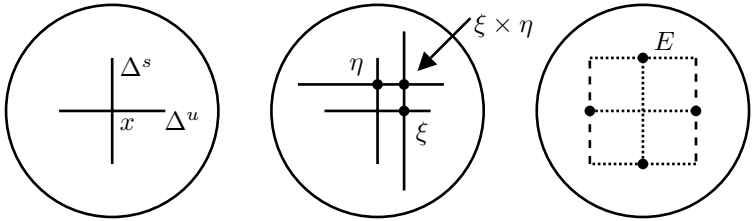
\0{\nota\it Fig.  1: The circles are a neighborhood of $x$ of size very
small compared to the curvature of the manifolds; the first picture
shows the axes; the intermediate picture shows the $\times$ operation
and $W^{u,\d}_\h, W^{s,\d}_\x$ (the horizontal and vertical segments
through $\h$ and $\x$, respectively, have size $\d$); the third picture
shows the rectangle $E$ with the axes and the four marked points are the
boundaries $\dpr\D^u$ and $\dpr\D^s$.  The picture refers to a two
dimensional case and the stable and unstable manifolds are drawn as
flat, \ie the $\D$'s is very small compared to the curvature of the
manifolds. Travsversality of $W^u_x,W^s_x$ is pictorially represented by
drawing the surfaces at $90^o$ angles, from \ref{GC}{2}{7}.\vfill}
\*
Consider now a partition $\EE=(E_1,\ldots,E_\NN)$
of $\CC$ into $\NN$ rectangles $E_j$ with pairwise disjoint interiors.
We call $\dpr_u\EE\=\cup_j\dpr_u E_j$ and $\dpr_s\EE\=\cup_j\dpr_s
E_j$: these are called respectively the {\it unstable boundary} of
$\EE$ and the {\it stable boundary} of $\EE$, or also the horizontal
and vertical boundaries of $\EE$, respectively.
 
We say that $\EE$ is a {\it Markov partition} if the transformation $S$
acting on the stable boundary of $\EE$ maps it into itself ($S
\dpr_s\EE\subset \dpr_s\EE$) and if, likewise, the map $S^{-1}$ acting on
the unstable boundary maps it into itself ($S^{-1}\dpr_u \EE\subset
\dpr_u\EE$).
 
The actual construction of the SRB distribution then proceeds from the
important geometric result of the theory of Anosov systems expressed by
what we shall call ``Sinai's first theorem'', \ref{S}{}{2}: \*
 
\0{\it Theorem: every transitive Anosov system admits a Markov partition}
$\EE$, {\it as fine as wished}.
\*
 
\0The term "fine" refers to the maximum size $\d_\EE$ of the sets $E\in\EE$
which can be fixed \ap to be smaller than any prefixed length ("as fine
as wished").
 
The theorem can be extended to imply the existence of more special
Markov partitions: for instance to show the existence of Markov
partitions with any one of the following three properties (the last
shows that the first two can be realized simultaneouly and will play a
key role in our analysis):
 
(1) The construction of $\EE$ can be done, \ref{Ga}{3}{18}, 
so that the horizontal
axes of $E_j$ {\it all lie on} $W^u_O$ (and the vertical on $W^s_O$) and
their union is a set that can be obtained from a single small connected
surface element $\lis\D$ of $W^u_O$ (resp. $\lis\D'$ of $W^s_O$)
containing $O$ by dilating it with a high iterate $S^Q$ of the time
evolution $S$. In other words the union $\cup_j\D^u_j$ of the horizontal
axes of the parallelograms $E_j\in\EE$ can be regarded as a single
connected surface which in turn is a good finite representation of the
attracting set.
 
Likewise the union of the stable axes can be regarded as a large
connected part of the stable manifold $W^s_O$.
 
(2) If the reversibility property holds it is clear that $i\EE$ is also a
a Markov partition. This follows from the definition of Markov
partition and from the fact that reversibility implies:
 
$$W^s_{x}=i W^u_{ix}\Eq(3.6)$$
 
\0(3) The definition of a Markov partition also implies that the
intersection of two Markov partitions is a Markov partition, hence it is
clear that there are Markov partitions $\EE$ that are reversible in the
sense that $\EE=i\EE$ (if this is not true one can just intersect $\EE$
and its $i$--image $i\EE$).
 
Once a Markov partition $\EE=(E_1,\ldots,E_\NN)$ is given one can
associate with each point $x\in \CC$ the {\it history} $\V
h(x)=(\ldots,h_{-1},h_0,h_1,\ldots)$ of the motion of $x$ on $\EE$: it
is the sequence of the names of the elements of $\EE$ that $x$ visits at
time $j$: $S^jx\in E_{h_j}$.
 
This is well defined for all $x\in\CC$ which do not visit in their
evolution the boundaries of the elements of $\EE$: the latter points
form a set of zero volume and can be disregarded for the purposes of our
analysis (having zero probability of being $\m_0$--randomly chosen).
The expansion and contraction properties of the sides of the
parallelograms imply that if the partition $\EE$ is {\it fine enough}
(precisely if the images under $S$ of each $E\in\EE$ intersects any
other $E'\in\EE$ in a connected set) they are small enough only one
point can have a given history and two points that have histories with
identical symbols $h_j$ for $|j|<N$ are close within a distance
$O(e^{-\l N})$.
 
The symbolic representation is "as good" as the decimal representation
of (the cartesian coordinates of) the points: provided we shall always
suppose the Markov partition to be ``fine enough'' (in the above sense)
it follows that the points $x\in\CC$ are determined by the digits of
their symbolic representation $\V h$ with an exponential precision, in
the sense that they get close at exponential rate in the number of
agreeing digits. It is however much better than the usual decimal
representations because it is {\it adapted} to the dynamics. The map $S$
in fact just becomes the translation of the the history $\V h(Sx)=\th \V
h$ if $\th$ is the left shift, obviously.
 
The above are properties that the history of $x$ on $\EE$ shares with
most fine partitions of phase space. There is however a property that
is enjoyed by the ordinary decimal representations of the cartesian
coordinates and that is \ap missing in the symbolic representations that
are derived from the histories on a generic partition of phase space.
 
Namely the ordinary representation of the cartesian coordinates by
decimal digits has the remarkable property that not only each point can
be represented symbolically but also {\it any sequence of digits}
determines a point.
 
The latter property does not hold in general for the
symbolic representation of a point via the history of its motion along a
partition. In fact if we define the {\it compatibility matrix} $M_{hh'}$
for the partition $\EE$ by setting $M_{hh'}=1$ if $SE^0_h\cap
E^0_{h'}\ne\emptyset$ (here $E^0$ denotes the interior of a set $E$) and
$M_{hh'}=0$ otherwise, then $\V h$ can be the history
of a point only if $M_{h_j h_{j+1}}\=1$. We call such sequences {\it
compatible} by nearest neighbours.
 
Note that if a system is a transitive Anosov system the matrix $M$ has
the {\it mixing property}, namely $M^n$ has strictly positive matrix
elements for all $n$ large enough.
 
But it is clear that the compatibility condition is not the only one:
one should define compatibility conditions linking three neighboring
times $j,j+1,j+2$ or four and more. This shows that constructing a
symbolic dynamics representation of a given dynamical system is a very
non trivial problem. The understanding of the compatibility conditions
essentially requires a complete solution of the dynamics in order to
tell \ap which are the symbolic sequences that can actually represent a
point, \ie it requires the construction of a {\it code} of the phase
space into a space of sequences.
 
What is special about Markov partitions is that we can \ap tell {\it
which are the possible histories} of points in $\CC$.  They are simply
{\it all} the sequences compatible by nearest neighbours: the space of
such sequences will be called $\KK_M$.  This is {\it very remarkable} as
in some sense it solves completely the problem of studying the motions of a
transitive Anosov system.
 
{\it Whatever the transitive Anosov system is the time evolution can be
represented simply as a shift on the space $\KK_M$ of biinfinite
sequences of symbols subject only to nearest neighbour compatibility
conditions} ("subshift of finite type" with a suitable compatibility
matrix).  Therefore the Anosov systems play the same role that the
harmonic oscillators play in perturbation theory: most systems are not
Anosov systems, but since we "fully" understand Anosov systems they
become the paradigm of non integrable motion just as the harmonic
oscillators are the paradigm of integrable motions, \ref{S}{}{2}.  \*
 
\0{\it \S4 Volume measure. Forward and backward SRB distributions.}
\numsec=4\numfor=1\*
 
An immediate consequence of the symbolic dynamics representation of the
points of $\CC$ in the cases in which $(\CC,S)$ is a transitive Anosov
system is that the Liouville measure (\ie the volume measure) on $\CC$
and its statistics are easily expressible in symbolic terms.
 
We consider the partition $\EE_T=\cap_{-T}^T S^{-j}\EE$ obtained by
intersecting the images under $S^j$, $j=-T,\ldots,T$ of $\EE$.  Then
$\EE_T$ is still a Markov partition and it is time reversal invariant
if $\EE$ is.
 
We note that the parallelograms of $\EE_T$ can be labeled by strings
of symbols $h_{-T},\ldots,h_T$ and they consist of the points $x$ such
that $S^k x\in E_{h_k}$ for $-T\le k\le T$.  In other words the
parallelograms consist of those points $x$ which in their time evolution
visit at time $k$ the parallelogram $h_k$ (one usually says that these
are the points whose {\it symbolic dynamics} on $\EE$ coincide, at the
times ("sites") $k$ between $-T$ and $T$).
 
In a parallelogram any point can be regarded as {\it center}.  However
there are special points that are usually taken as centers because they
play a special role.  Accordingly we shall take as center
$x_{h_{-T},\ldots,h_T}$ of a parallelogram $E_{h_{-T},\ldots,h_T}$ a
point whose symbolic dynamics string $\V j$ at the times $k$ with
$|k|>T$ is fixed in a standard way; \ie by defining $h_k$ for $k>T$
(respectively $k<-T$) as a compatible sequence depending only on $h_T$
(respectively $h_{-T}$).  We cannot in general make the simple choice
of continuing ${h_{-T},\ldots,h_T}$ at times $k>T$ or $k<-T$ with a
fixed symbol because this may lead to an incompatible sequence (hence
the choice that we make is ``the simplest'' possible: it leaves some
arbitrariness because there are many strings of symbols that start with
a given symbol.  The arbitrariness is however irrelevant for what
follows).
 
The possibility of the above continuation of finite strings into
infinite strings relies on the mixing property of the matrix $M$,
consequence of the transitivity of $(\CC,S)$.
 
Another simple choice of the center is to pick a periodic point in
$E_{h_{-T},\ldots,h_T}$: this can be easily done by continuing  the
string $h_{-T},\ldots,h_T$ beyond $T$ in a standard way (in the above
sense) to $h_{-T-n},\ldots,h_{T+n}$ so that $M_{h_{T+n}h_{-T-n}}=1$ and
then continuing it {\it periodically}.
 
We can now define probability distributions on the phase space $\CC$
simply by defining probability distributions on the  space $\KK_M$
of compatible sequences with compatibility matrix $M$ of a fine Markov
partition $\EE$.
 
The simplest way of constructing probability distributions on $\KK_M$ is
to think of the symbolic sequences as {\it spin} chains on a one
dimensional lattice (the time in our case) and to assign an energy to
each chain and construct the corresponding Gibbs probability
distribution.
 
We recall that an energy function for a chain of spins is a function
$\ell_0(\V h)$, to be called the energy of interaction of the spin at
$0$ with the other spins of the chain, which has {\it short range}, \eg
for some $k,\k>0$ is is $|\ell_0(\V h)-\ell_0(\V h')|\le k e^{-\k |N|}$
if $\V h$ and $\V h'$ agree on the sites $j$ with $|j|<N$.
 
In other words an energy function has short range if it depends on a
finite number of spins ("Ising chains") or if it depends
{\it"exponentially little"} on the "far" spins.

Then one defines a Gibbs distribution with energy functions $\{\ell_j(\V
h)\}$ by setting:
 
$$\ig \tilde\m(d\V h') \tilde F(\V h')=
\lim_{T\to\io}\fra{\sum_{h_{-T}\ldots,h_T} \tilde F(\V h)
e^{-\sum_{j=-T}^{T-1}
\ell_j(S^j\V h)}}{\sum_{h_{-T}\ldots,h_T} e^{-\sum_{j=-T}^{T-1}
\ell_j(S^j\V h)}}\=\lim_{T\to\io}
\tilde\m_T(\tilde F)\Eq(4.1)$$
here the infinite sequence $\V h$ is a "standard" (in the above sense)
continuation of the string $h_{-T},\ldots,h_{T}$ to an infinite
compatible sequence.
 
If $\ell_j$ verifies mild translation invariance properties, \eg if
$\ell_j(\V h)=\ell_+(\V h)$ for all $j\ge0$ and $\ell_j(\V h)=\ell_-(\V h)$
for all $j<0$, the limit exits and is unique.
 
The metric dynamical system $(\KK_M,\th,\tilde\m)$ has very strong
cluster properties, for instance it is exponentially mixing
in the sense that, if $\media{\cdot}_\sim$ denotes $\tilde\m$--average
the correlation:
 
$$\media{\tilde F(S^j\cdot)\tilde F(\cdot)}_\sim-
\media{\tilde F(S^j\cdot)}_\sim\media{\tilde
F(\cdot)}_\sim\tende{j\to\pm\io}0\Eq(4.2)$$
and the convergence to zero is exponentially fast, for all functions on
$\KK_M$ with short range in the above sense.
 
One can construct in this way three gibbsian probability distributions
that we call $\tilde\m_0,\tilde \m_+,\tilde \m_-$. If $\V h\otto x(\V
h)$ is the code that associates with each compatible history $\V h\in
\KK_M$ the point $x\in \CC$ with that history, the distributions
$\tilde\m,\tilde\m_\pm$ are the Gibbs distributions with respective
potentials:
 
$$\eqalign{
\ell_j(\V h)=&\cases{\l_+(\V h)=\log\L_u(x(\V h))& if $j>0$\cr
\l_-(\V h)=-\log\L_s(x(\V h))& if $j<0$\cr}\cr
\ell_j(\V h)=&\l_+(\V h)\cr
\ell_j(\V h)=&\l_-(\V h)\cr}\Eq(4.3)$$
Since the map $\V h\to x(\V h)$ is such that $d(x(\V h),x(\V h'))< B
e^{-\l N}$ if the histories $\V h,\V h'$ agree between time $-N$ and
time $N$, and since the functions $\L_u(x),\L_s(x)$ are H\"older
continuous (see \S1), it follows immediately that $\ell_j(\V h)$ have
short memory and therefore they define Gibbs states $\tilde \m_0, \tilde
\m_+, \tilde \m_-$.
 
The above distributions $\tilde \m$ can be coded back to distributions
on $\CC$ simply by using the code $\V h\to x(\V h)$ between $\KK_M$ and
$\CC$ and setting: $\m(x(G))=\tilde \m(G)$ for a generic set $G$.
 
The following theorem by Sinai is the fundamental result, \ref{S}{}{2}:
\*
 
\0{\it Theorem 2: Coding back into distributions on $\CC$ the
distributions $\tilde\m_0,\tilde\m_\pm$ one obtains three distributions
$\m_0,\m_+,\m_-$. The first $\m_0$ is proportional to the volume, and in
general it is not $S$--invariant, while the two others are the forward
and backward statistics of the volume.}
\*
 
Thus this theorem solves completely the problem of the existence of the
SRB distributions $\m)\pm$. It also gives us a concrete and usable
representation for the SRB distribution via \equ(4.3) and \equ(4.1).
 
The choice of the ``standard'' continuation of $h_{-T},\ldots,h_{T}$ to
$\V h\in K_M$ in \equ(4.1) is quite arbitrary: but the above theorem
implies that it is immaterial. In particular a very popular choice is
the periodic continuation which leads to the alternative representation
of the SRB distributions $\m_+$ and $\m_-$ known as the {\it periodic
orbit expansion}. Although usually (and inexplicably) surrounded by an
aura of mystery this is a representation that is easier to communicate
to unfavourable audiences as it does not require the symbolic dynamics
analysis ({\it on which, however, it rests}):
 
$$\ig F(x)\m_+(dx)=\lim_{T\to\io}
\fra{\sum_{S^{2T+1}x=x} F(x) e^{-\sum_{j=-T}^{T-1}\l_+(S^j x)}}
{\sum_{S^{2T+1}x=x} e^{-\sum_{j=-T}^{T-1}\l_+(S^j x)}}\Eq(4.4)$$
which is a formula that is an immediate (non trivial) consequence (if
one takes for granted the theory of one dimensional Gibbs states) of the
previous
\equ(4.1),\equ(4.3).
 
The periodic orbit representation of measures on on $\CC$ has the
disadvantage that the volume distribution $\m_0$ does not have a natural
representation: only $\m_+$ (and $\m_-$) are easily represented. This is
a drawback as \equ(4.1),\equ(4.3) make very transparent the relationship
between $\m_0,\m_\pm$ and makes it obvious why $\m_\pm$ are the forward
and backward statistics of $\m_0$.
 
It is also possible to write \equ(4.1) with $\ell_j=\l_+$ "directly" as
a distribution on $\CC$ as:
 
$$\ig\m_+(dx) F(x)=\lim_{T\to\io}\m_T(F)=
\lim_{T\to\io}
\fra{\sum_j e^{-\sum_{k=-T}^{T-1}\l_+(x_j)(S^kx_j)} F(x_j)}
{\sum_j e^{-\sum_{k=-T}^{T-1}\l_+(x_j)(S^kx_j)}}\Eq(4.5)$$
where the sum runs over the elements $E_j$ of the pavement $S^{T}\EE\vee
\ldots\vee S^{-T}E_{h_{T}}$, $\V h$ is the extension
of $(h_{-T},\ldots,h_{T})$ to a string in $\KK_M$ and $x_j\=x(\V h)$.
The function $F$ is any smooth (\ie H\"older continuous) function on
$\CC$.
 
The relation \equ(4.5) is also written as:
 
$$\ig\m_+(dx) F(x)= \lim_{T\to\io} \fra{\sum_j \L_{u,2T}(x_j)^{-1}
F(x_j)}{\sum_j\L_{u,2T}(x_j)^{-1}}\=\lim_{T\to\io}\m_T(F)\Eq(4.6)$$
because by definition the jacobian determinant $\L_{u,t}(x)$ of the
jacobian of the map $S^t$ as a map of $W^u_{S^{-t/2}x}$ to
$W^u_{S^{t/2}x}$ is, see \equ(3.1),
$\exp{\sum_{k=-t/2}^{t/2-1}\l_+(x_j)(S^kx_j)}$.
 
We now show informally why \equ(4.6) implies the fluctuation theorem,
\ref{GC}{1}{13},\ref{GC}{2}{7},\ref{Ga}{2}{17}.
 
We first evaluate the probability,  with respect to $\m_{\t/2}$ of
\equ(4.6),
of $\s_\t(x)/\media{\s}_+\in I_p$ divided by
the probability (with respect to the same distribution)
that
$\s_\t(x)/\media{\s}_+\in I_{-p}$. This is:
$$\fra{\sum_{j,\,a_\t(x_j)=p} \lis\L_{u,\t}^{-1}(x_j)}
{\sum_{j,\, a_\t(x_j)=-p} \lis\L_{u,\t}^{-1}(x_j)}\Eq(4.7)$$
Since $\m_{\t/2}$ in \equ(4.6) is only an approximation to $\m_+$
{\it an error is involved in using} \equ(4.7) as a formula for the same
ratio computed by using the true $\m_+$ instead of $\m_{\t/2}$.
 
It can be shown that this error can be estimated to affect the result
only by a factor bounded above and below uniformly in $\t,p$,
\ref{GC}{1}{13}{2}{}. This is a remark technically based on the
thermodynamic analogy pointed out in \equ(4.1) above.
 
We now try to establish a one to one correspondence between the addends
in the numerator of \equ(4.7) and the ones in the denominator,
aiming at showing that corresponding addends have a {\it constant
ratio} which will, therefore, be the value of the ratio in \equ(4.7).
 
This is possible because of the reversibility property: it will be used
in the form of its consequences given by the relations \equ(3.2).
 
The ratio \equ(4.7) can therefore be written simply as:
$$\fra{\sum_{E_j, a_\t(x_j)=p} \lis\L_{u,\t}^{-1}(x_j)}
{\sum_{E_j, a_\t(x_j)=-p} \lis\L_{u,\t}^{-1}(x_j)}
\=\fra{\sum_{E_j, a_\t(x_j)=p} \lis\L_{u,\t}^{-1}(x_j) }
{\sum_{E_j, a_\t(x_j)=p} \lis\L_{s,\t}(x_j)}\Eq(4.8)$$
where $x_j\in E_j$ is the center in $E_j$. In deducing the second
relation we make use of the existence of the time reversal symmetry $i$
and of \equ(3.5), and assume that the centers $x_{j}, x_{j'}$ of $E_j$
and $E_{j'}=iE_j$ are chosen so that $x_{j'}=ix_j$.
 
It follows then that the ratios between corresponding terms in the ratio
\equ(4.8) is equal to $\lis\L_{u,\t}^{-1}(x)\lis\L_{s,\t}^{-1}(x)$.
This differs from the reciprocal of the total change of phase space
volume over the $\t$ time steps between the point $S^{-\t/2}x$ and
$S^{\t/2 }x$ only because it does not take into account the ratio of
the sines of the angles $a(S^{-\t/2}x)$ and $a(S^{\t/2}x)$ formed
by the stable and unstable manifolds at the points $S^{-\t/2}x$ and
$S^{\t/2}x$, see footnote ${}^1$.  But $\lis\L_{u,\t}^{-1}(x)
\lis\L_{s,\t}^{-1}(x)$ will differ from the
actual phase space contraction under the action of $S^\t$, as a map
between $S^{-\t/2}x$ and $S^{\t/2}x$, by a factor that can be bounded
between $B^{-1}$ and $B$ with
$B=\max_{x,x'}\fra{|\sin a(x)|}{|\sin a(x`)|}$ which is finite by the
transversality of the stable and unstable manifolds.
 
Now for all the points $x_j$ in \equ(4.8), the reciprocal of the total
phase space volume change over a time $\t t_0$ is
$e^{\s_\t(x_j)\media{\s}_+\t}$, which (by the constraint imposed on the
summation labels $\s_\t/\media{\s}_+=p$)  equals
$e^{\t\media{\s}_+\,p}$. Hence the ratio \equ(4.7) will be
$e^{\t\media{\s}_+\,p}$.  It is important to note that there are two
errors ignored here, as pointed out in the discussion
above. They imply that the argument of the exponential
{\it is correct up to $p,\t$ independent corrections} (which are in fact
observed in the experiment as fig.3 of \ref{ECM}{2}{19} shows, or as it is shown
by Fig. 2 below).  One should note that other errors may arise because
of the approximate validity of our main chaotic assumption (which states
that things go "as if" the system was Anosov): they may depend on $N$
and we do not control them except for the fact that, if present, their
relative value should tend to $0$ as $N\to\io$: there may be (and very
likely there are) cases in which the chaotic hypotesis is not reasonable
for small $N$ (\eg systmes like the Fermi-Pasta-Ulam chains) but it
might be correct for large $N$.  We also mention that for some systems
with small $N$ for which the chaotic hypothesis may be already regarded
as valid (\eg model 1 with $N=1$ in \ref{CELS}{}{20}).
 
The $p$ independence of the ratio \equ(3.5) is therefore a key test of
the theory (and it should hold with corrections of order $O(\t^{-1})$ if
$\z(p)$ is evaluated by its finite $\t$ approximation $\z_\t(p)$).
 
The fluctuation theorem is not immediately applicable in many cases.
For instance in the cases in which the attractor is an axiom A attractor
smaller than the whole phase space.
 
If however we imagine that the chaotic hypothesis holds in the stronger
sense that the system verifies the Axiom C then we can use that
the map $\tilde\imath$ that maps the backward attractor $\AA_-$
into the forward attractor $\AA_+$ and combine it with the time reversal
symmetry $i$ to generate a map $i^*=\tilde\imath i$ which maps the two
poles $\AA_+$ and $\AA_-$ into themselves (see \S2). The new map $i^*$
is {\it a time reversal symmetry} that anticommutes with the time
evolution, \ie $Si^*=i^* S^{-1}$ {\it leaving $\AA_+$ invariant}.
 
Thus the fluctuation therem holds for such systems because we can think
that $(\AA_+,S)$ is a reversible transitive Anosov system to which we
can apply the above argunents.
 
Of course if we take the above viewpoint the phase space contraction
that the theorem referes to should {\it no longer} be the contraction of
the total volume but {\it just} the contraction of the surface measure
of $\AA_+$ which will now play the role of $\m_0$.
 
The latter result might be difficult to use as the attractor $\AA_+$ is
very likely to be an unreachable object. And in fact to be applied new
properties must hold. It is remarkable that such poperties had already
been observed in some special systems, much earlier than the fluctuation
theorem. We discuss the applications of the theorem in \S4.
 
Note that the theorem is a {\it large fluctuation} theorem: therefore it
is very hard to test. The measure of how large are the fluctuations can
be grasped also by the attempts that have been proposed in the
literature to link the phenomena that occurr in models like the first
one in \S2 and the {\it Loschmidt paradox}, \ref{HHP}{}{21}.  Clearly the
deviations of $p$ from the average are large deviations that, in order
to be neasured, certainly involve observations of currents opposite to
the field and violating the second law.
 
%\ifnum\mgnf=0\vglue3cm\fi
%\kern4cm
\*
 
\0{\it\S5. Applications: reversible conduction, fluids. Onsager
reciprocity. Dynamical ensembles. Irreversible systems.}
\numsec=5\numfor=1\*
 
The first applications are numerical tests. The basic result is
\ref{ECM}{2}{19} which was the origin of the whole story: the chaotic
hypothesis was formulated to find a theoretical explanation of the
results of this important experiment. The latter authomatically provides
a first test of the hypothesis.
 
Experiments designed to test the hypthesis have since been performed.  I
quote here the first of them, \ref{BGG}{}{14}, that studies the reversible
conduction model in example 1, \S2.  The following graphs are the
results of tests of the fluctuation theorem in a $2$ or $10$
particles system of hard spheres enclosed in a periodic box containing
two hard sphere fixed obstacles and subject to a {\it very} large
electric field ($E=1$ which is huge in physical units if one thinks of
the particles as electrons in a crystal).
%\1
 
The following are the graphs of the empirically determined
$x(p)=(\z(-p)-\z(p))/\media{\s}_+$. The case of $2$ particles and
semiperiodic boundary conditions yields:

\dimen2=25pt \advance\dimen2 by 95pt
\dimen3=103pt \advance\dimen3 by 40pt
\eqfigbis{\dimen3}{125pt}{
\ins{123pt}{0pt}{$\hbox to20pt{\hfill ${\scriptstyle p}$
\hfill}$}
\ins{0pt}{\dimen2}{$\hbox to25pt{\hfill ${\scriptstyle x(p)}$
\hfill}$}
\ins{71pt}{113pt}{$\hbox to51pt{\hfill ${\scriptstyle \tau=20}$
\hfill}$}
\ins{20pt}{3pt}{$\hbox to 15pt{\hfill${\scriptstyle 0}$
\hfill}$}
\ins{35pt}{3pt}{$\hbox to 15pt{\hfill${\scriptstyle 1}$
\hfill}$}
\ins{51pt}{3pt}{$\hbox to 15pt{\hfill${\scriptstyle 2}$
\hfill}$}
\ins{66pt}{3pt}{$\hbox to 15pt{\hfill${\scriptstyle 3}$
\hfill}$}
\ins{81pt}{3pt}{$\hbox to 15pt{\hfill${\scriptstyle 4}$
\hfill}$}
\ins{97pt}{3pt}{$\hbox to 15pt{\hfill${\scriptstyle 5}$
\hfill}$}
\ins{112pt}{3pt}{$\hbox to 15pt{\hfill${\scriptstyle 6}$
\hfill}$}
\ins{0pt}{21pt}{$\hbox to 20pt{\hfill${\scriptstyle 0}$
\hfill}$}
\ins{0pt}{31pt}{$\hbox to 20pt{\hfill${\scriptstyle 1}$
\hfill}$}
\ins{0pt}{41pt}{$\hbox to 20pt{\hfill${\scriptstyle 2}$
\hfill}$}
\ins{0pt}{50pt}{$\hbox to 20pt{\hfill${\scriptstyle 3}$
\hfill}$}
\ins{0pt}{60pt}{$\hbox to 20pt{\hfill${\scriptstyle 4}$
\hfill}$}
\ins{0pt}{70pt}{$\hbox to 20pt{\hfill${\scriptstyle 5}$
\hfill}$}
\ins{0pt}{80pt}{$\hbox to 20pt{\hfill${\scriptstyle 6}$
\hfill}$}
\ins{0pt}{89pt}{$\hbox to 20pt{\hfill${\scriptstyle 7}$
\hfill}$}
\ins{0pt}{99pt}{$\hbox to 20pt{\hfill${\scriptstyle 8}$
\hfill}$}
}{figbr20}{\ins{123pt}{0pt}{$\hbox to20pt{\hfill ${\scriptstyle p}$
\hfill}$}
\ins{0pt}{\dimen2}{$\hbox to25pt{\hfill ${\scriptstyle x(p)}$
\hfill}$}
\ins{71pt}{113pt}{$\hbox to51pt{\hfill ${\scriptstyle \tau=100}$
\hfill}$}
\ins{13pt}{3pt}{$\hbox to 25pt{\hfill${\scriptstyle 0.0}$
\hfill}$}
\ins{38pt}{3pt}{$\hbox to 25pt{\hfill${\scriptstyle 0.5}$
\hfill}$}
\ins{64pt}{3pt}{$\hbox to 25pt{\hfill${\scriptstyle 1.0}$
\hfill}$}
\ins{89pt}{3pt}{$\hbox to 25pt{\hfill${\scriptstyle 1.5}$
\hfill}$}
\ins{0pt}{20pt}{$\hbox to 20pt{\hfill${\scriptstyle 0}$
\hfill}$}
\ins{0pt}{53pt}{$\hbox to 20pt{\hfill${\scriptstyle 1}$
\hfill}$}
\ins{0pt}{85pt}{$\hbox to 20pt{\hfill${\scriptstyle 2}$
\hfill}$}
}{figbr21}{}
\*
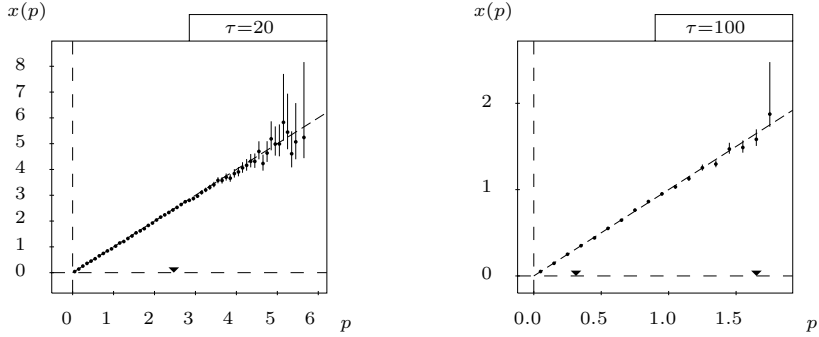
\didascalia{Fig.1: The graphs for the fluctuation theorem test
$N=2$; the dashed line is the fluctuation theorem
prediction, $\t=20,100$. The arrows mark the point at
distance $\sqrt{\media{(p-1)^2}}$ from $1$, from \ref{BGG}{}{14}.}
 
And the case of $10$ particles and semiperiodic boundary conditions
gives:
 
\dimen2=25pt \advance\dimen2 by 95pt
\dimen3=103pt \advance\dimen3 by 40pt
\dimenfor{\dimen3}{125pt}
\eqfigfor{
\ins{123pt}{0pt}{$\hbox to20pt{\hfill ${\scriptstyle p}$
\hfill}$}
\ins{0pt}{\dimen2}{$\hbox to25pt{\hfill ${\scriptstyle x(p)}$
\hfill}$}
\ins{71pt}{113pt}{$\hbox to51pt{\hfill ${\scriptstyle \tau=20}$
\hfill}$}
\ins{18pt}{3pt}{$\hbox to 19pt{\hfill${\scriptstyle 0}$
\hfill}$}
\ins{37pt}{3pt}{$\hbox to 19pt{\hfill${\scriptstyle 1}$
\hfill}$}
\ins{56pt}{3pt}{$\hbox to 19pt{\hfill${\scriptstyle 2}$
\hfill}$}
\ins{75pt}{3pt}{$\hbox to 19pt{\hfill${\scriptstyle 3}$
\hfill}$}
\ins{94pt}{3pt}{$\hbox to 19pt{\hfill${\scriptstyle 4}$
\hfill}$}
\ins{0pt}{21pt}{$\hbox to 20pt{\hfill${\scriptstyle 0}$
\hfill}$}
\ins{0pt}{34pt}{$\hbox to 20pt{\hfill${\scriptstyle 1}$
\hfill}$}
\ins{0pt}{48pt}{$\hbox to 20pt{\hfill${\scriptstyle 2}$
\hfill}$}
\ins{0pt}{61pt}{$\hbox to 20pt{\hfill${\scriptstyle 3}$
\hfill}$}
\ins{0pt}{75pt}{$\hbox to 20pt{\hfill${\scriptstyle 4}$
\hfill}$}
\ins{0pt}{88pt}{$\hbox to 20pt{\hfill${\scriptstyle 5}$
\hfill}$}
\ins{0pt}{102pt}{$\hbox to 20pt{\hfill${\scriptstyle 6}$
\hfill}$}
}{figbr100}{\ins{123pt}{0pt}{$\hbox to20pt{\hfill ${\scriptstyle p}$
\hfill}$}
\ins{0pt}{\dimen2}{$\hbox to25pt{\hfill ${\scriptstyle x(p)}$
\hfill}$}
\ins{71pt}{113pt}{$\hbox to51pt{\hfill ${\scriptstyle \tau=40}$
\hfill}$}
\ins{19pt}{3pt}{$\hbox to 15pt{\hfill${\scriptstyle 0.0}$
\hfill}$}
\ins{35pt}{3pt}{$\hbox to 15pt{\hfill${\scriptstyle 0.5}$
\hfill}$}
\ins{51pt}{3pt}{$\hbox to 15pt{\hfill${\scriptstyle 1.0}$
\hfill}$}
\ins{67pt}{3pt}{$\hbox to 15pt{\hfill${\scriptstyle 1.5}$
\hfill}$}
\ins{83pt}{3pt}{$\hbox to 15pt{\hfill${\scriptstyle 2.0}$
\hfill}$}
\ins{99pt}{3pt}{$\hbox to 15pt{\hfill${\scriptstyle 2.5}$
\hfill}$}
\ins{115pt}{3pt}{$\hbox to 15pt{\hfill${\scriptstyle 3.0}$
\hfill}$}
\ins{0pt}{20pt}{$\hbox to 20pt{\hfill${\scriptstyle 0}$
\hfill}$}
\ins{0pt}{44pt}{$\hbox to 20pt{\hfill${\scriptstyle 1}$
\hfill}$}
\ins{0pt}{68pt}{$\hbox to 20pt{\hfill${\scriptstyle 2}$
\hfill}$}
\ins{0pt}{93pt}{$\hbox to 20pt{\hfill${\scriptstyle 3}$
\hfill}$}
}{figbr101}{\ins{123pt}{0pt}{$\hbox to20pt{\hfill ${\scriptstyle p}$
\hfill}$}
\ins{0pt}{\dimen2}{$\hbox to25pt{\hfill ${\scriptstyle x(p)}$
\hfill}$}
\ins{71pt}{113pt}{$\hbox to51pt{\hfill ${\scriptstyle \tau=60}$
\hfill}$}
\ins{15pt}{3pt}{$\hbox to 21pt{\hfill${\scriptstyle 0.0}$
\hfill}$}
\ins{37pt}{3pt}{$\hbox to 21pt{\hfill${\scriptstyle 0.5}$
\hfill}$}
\ins{58pt}{3pt}{$\hbox to 21pt{\hfill${\scriptstyle 1.0}$
\hfill}$}
\ins{80pt}{3pt}{$\hbox to 21pt{\hfill${\scriptstyle 1.5}$
\hfill}$}
\ins{102pt}{3pt}{$\hbox to 21pt{\hfill${\scriptstyle 2.0}$
\hfill}$}
\ins{0pt}{20pt}{$\hbox to 20pt{\hfill${\scriptstyle 0.0}$
\hfill}$}
\ins{0pt}{35pt}{$\hbox to 20pt{\hfill${\scriptstyle 0.5}$
\hfill}$}
\ins{0pt}{50pt}{$\hbox to 20pt{\hfill${\scriptstyle 1.0}$
\hfill}$}
\ins{0pt}{65pt}{$\hbox to 20pt{\hfill${\scriptstyle 1.5}$
\hfill}$}
\ins{0pt}{80pt}{$\hbox to 20pt{\hfill${\scriptstyle 2.0}$
\hfill}$}
\ins{0pt}{94pt}{$\hbox to 20pt{\hfill${\scriptstyle 2.5}$
\hfill}$}
}{figbr102}{\ins{123pt}{0pt}{$\hbox to20pt{\hfill ${\scriptstyle p}$
\hfill}$}
\ins{0pt}{\dimen2}{$\hbox to25pt{\hfill ${\scriptstyle x(p)}$
\hfill}$}
\ins{71pt}{113pt}{$\hbox to51pt{\hfill ${\scriptstyle \tau=100}$
\hfill}$}
\ins{8pt}{3pt}{$\hbox to 33pt{\hfill${\scriptstyle 0.0}$
\hfill}$}
\ins{41pt}{3pt}{$\hbox to 33pt{\hfill${\scriptstyle 0.5}$
\hfill}$}
\ins{75pt}{3pt}{$\hbox to 33pt{\hfill${\scriptstyle 1.0}$
\hfill}$}
\ins{0pt}{19pt}{$\hbox to 20pt{\hfill${\scriptstyle 0.0}$
\hfill}$}
\ins{0pt}{42pt}{$\hbox to 20pt{\hfill${\scriptstyle 0.5}$
\hfill}$}
\ins{0pt}{65pt}{$\hbox to 20pt{\hfill${\scriptstyle 1.0}$
\hfill}$}
\ins{0pt}{89pt}{$\hbox to 20pt{\hfill${\scriptstyle 1.5}$
\hfill}$}
}{figbr103}{}
 
%%%%%%%%%%%%%%%%%%%%%%%
 
\*
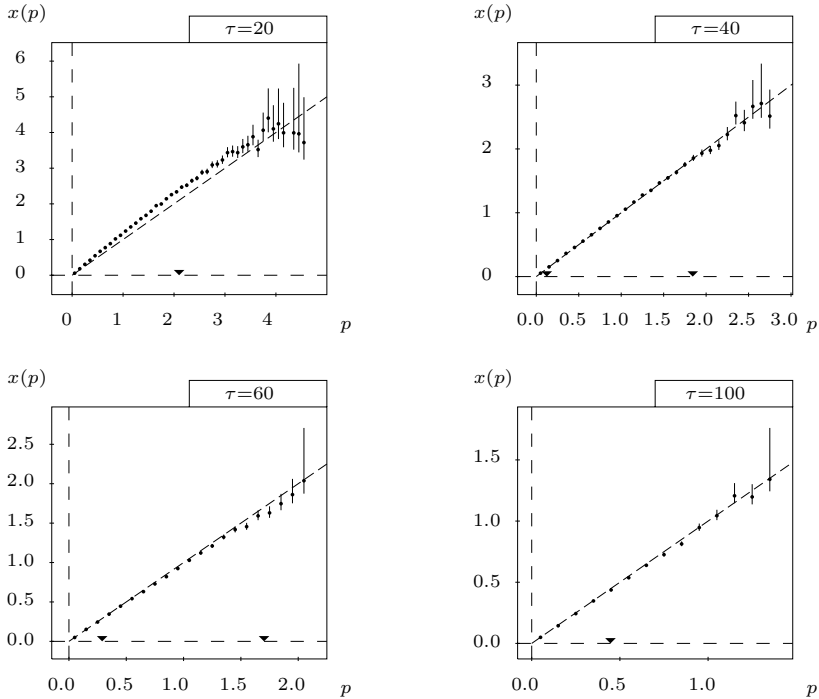
\didascalia{Fig.2: The linear fluctuation test, $\t=20,40,80,100$ for
$N=10$. The
dashed line is the fluctuation theorem prediction for $\t=+\io$. The
arrows mark the point at distance $\sqrt{\media{(p-1)^2}}$ from $1$,
from \ref{BGG}{}{14}.}
 
According to the theorem the graphs should be a straight line with slope
$1$.   The graphs show the data as well as the theoretical line with
slope $1$ (dashed).  The function $\z(p)$, hence $x(p)$,  in fact
depends on $\t$, the time interval over which the fluctuations are
observed and should be more properly denoted $\z_\t(p)$.  We show the
graphs of $x(p)$ versus $p$ for various $\t$: the slope $1$ is exact
only for $\t=+\io$ but we see that it can already be observed for
reasonable $\t$--values.

The experiments in \ref{BGG}{}{14} showed various other interesting
phenomena: namely the fact that for large fields the attractor is
certainly smaller than the full phase space.
 
The interpretation of the fact that in spite of that the experimental
results still matched with \equ(3.5), generated the idea of
strengthening of the chaotic hypothesis by replacing the assumption
that the system has attractors that can be regaded as Anosov systems by
the stronger assumption that the system verifies Axiom C, which should
have a general validity in time reversible systems: see \ref{BG}{}{6}.
 
If $(\CC,S)$ is reversible and verifies Axiom C the map $\tilde\imath$
defined via geometric considerations in \S2 will commute with $i$ (by
the covariance under $i$ of the manifolds used to build it) and
therefore the map $\tilde\imath i=i^*$ will leave invariant the poles
$\O_\pm$ of the system and on them it will {\it anticommute} with $S$.
 
This means that even though time reversal symmetry
is "lost" or "broken" on $\O_\pm$ there is {\it another} transformation
($i^*$) leaving $\O_\pm$ invariant and anticommuting with $S$, \ie
which plays the role of a time reversal symmetry for the restriction of
$S$ to $\O_+$ (and $\O_-$).
 
This implies that in reversible Axiom C systems {\it time reversal is
undestructible}: if the  attractor becomes smaller than the full phase
space one can say that the  time reversal symmetry is {\it spontaneously
broken}. But on the  attractor one can still define a map $i^*\ne i$
which anticommutes with  time and leave the attractor invariant. This is
the "real" time reversal  for the attractor, or the {\it local time
reversal}, see \ref{BG}{}{6}: while the "true" time reversal appears as
a broken symmetry to an observer that only observes motions that are
attracted by $\O_+$.
 
A situation reminiscent of the violation of time reversal in elemntary
particles physics, where the time reversal symmetry, or T symmetry, is
broken but there is another symmetry that chab=nges the sign to time,
the TCP symmetry. Here $i$ plays the role of $T$ and $i^*$ the role of
$TCP$ (and $\tilde \imath $ that of CP).
 
Hence the volume on the attractor will contract obeying the
fluctutation theorem.  This in itself is not sufficient to study the
phase space volume fluctuations because, as said above, one does not
have direct access to the surface area of the attractor (and to the
attractor itself, as a matter of fact).
 
Then manifestly one would be back with an Anosov system (on a lower
dimensional manifold) and a version of the fluctuation theorem would
still hold.  Furthermore one could say that this is only a different
interpretation of the chaotic principle.
 
If this picture is correct we can write the phase space contraction rate
(see \equ(3.2) $\s(x)=\s_0(x)+\s_\perp(x)$ where $\s_0(x)$ is
the contraction rate on the surface on which the attractor lies and
$\s_\perp(x)$ is the contraction rate of the part of the stable manifold
of the attractor which is not on the attractor itself (the angle between
the part of the stable manifold sticking out of the attractor and the
attractor itself is disregarded here as we think that it is bounded away
from $0$ and $\p$ since the attractor is compact).
 
Local time reversal will change the sign {\it only of} $\s_0(x)$ and the
fluctuation theorem should apply to the fluctuations of $\s_0$.  But
$\s_0(x)$ is {\it not} directly accessible to measurement: nevertheless
we can still study its fluctuations via the following {\it heuristic}
analysis, \ref{BGG}{}{14}.
 
Considerable help comes from a remarkable theorem that was discovered
experimentally in \ref{EM}{}{22}, \ref{ECM}{1}{23}.  In the full phase
space of the equations in the reversible example 1) in \S1 the Lyapunov
exponents verify a {\it pairing rule}.  Namely if $2D=4N-2$ is the number
of exponents and the first $D=2N-1$ exponents,
$\l^+_1,\ldots,\l^+_{2N-1}$, are ordered in decreasing order and the
next $2N-1$, $\l^-_1,\ldots,\l^-_{2N-1}$, are ordered in increasing
order then:
 
$$\fra{\l^+_j+\l^-_j}2= const\qquad {\rm for\ all}\
j=1,\ldots,D\Eq(5.1)$$
The constant will be called "{\it pairing level}" or "{\it pairing
constant}": it must be $\fra1{2D}\media{\s}_+$.
 
The pairing rule, in fact, formally holds in the present example
as a consequence of \ref{DM}{}{24}.
 
In the cases in which \equ(5.1) has been proved, \ref{DM}{}{24}, it holds
also in a far {\it stronger} sense: the {\it local Lyapunov exponents},
of which the Lyapunov exponents are the averages, are paired as in
\equ(3.2) to a constant that is $j$ independent but, of course, is
dependent on the point in phase space.  We call this the {\it strong
pairing rule}.  From the proof in \ref{DM}{}{24} of the pairing rule one
sees that the jacobian matrix $J=\dpr S$ of the map $S$ is such that
$\sqrt{J^*J}$ has $D$ pairs of eigenvalues ($D=2N-1$ in the case of
example 1, \S2) and the logarithms of each pair add up to $\s(x)$ (see
also \equ(2.4),\equ(2.5)).
 
The simplest interpretation of this, consistent with  the proposed
attractor picture, is that pairs with elements of opposite
signs describe expansion {\it on the manifold} on which the attractor
lies.  While the $M\le D$ pairs consisting of two negative exponents
describe contraction of phase space {\it
transversally to the manifold} on which the attractor lies.
 
Then we would have $\s_0(x)=(D-M)\s(x)$ and we should have a fluctuation
law for the quantity $p(x)$ associated with $\s_0(x)$ defined by
\equ(2.6) and \equ(2.7) with $\s_0$ replacing $\s$, \ie (accepting the
above heuristic argument, taken from \ref{BGG}{}{14}):
$$\s_{0\t}(x)=\fra{(D-M)}D\media{\s}_+ p(x)\Eq(5.2)$$
\ie a law identical to \equ(2.10) {\it up to a correcting factor}
$1-\fra{M}D$:
$$\fra1{p\t\media{\s}_+}
\log\fra{\p_\t(p)}{\p_\t(-p)}=(1-\fra{M}D)\,p\Eq(5.3)$$
The graphs of Fig.2 are relative to an experiment in which we see that
there may be one negative exponents in excess over the positive ones as
said above.
 
\dimen2=25pt \advance\dimen2 by 198pt
\dimen3=276pt \advance\dimen3 by 70pt
\eqfig{\dimen3}{223pt}{
\ins{306pt}{0pt}{$\hbox to20pt{\hfill ${\scriptstyle j}$\hfill}$}
\ins{24pt}{\dimen2}{$\hbox to25pt{\hfill ${\scriptstyle \lambda}$\hfill}$}
\ins{33pt}{3pt}{$\hbox to 12pt{\hfill${\scriptstyle 0}$\hfill}$}
\ins{46pt}{3pt}{$\hbox to 12pt{\hfill${\scriptstyle 1}$\hfill}$}
\ins{59pt}{3pt}{$\hbox to 12pt{\hfill${\scriptstyle 2}$\hfill}$}
\ins{72pt}{3pt}{$\hbox to 12pt{\hfill${\scriptstyle 3}$\hfill}$}
\ins{85pt}{3pt}{$\hbox to 12pt{\hfill${\scriptstyle 4}$\hfill}$}
\ins{97pt}{3pt}{$\hbox to 12pt{\hfill${\scriptstyle 5}$\hfill}$}
\ins{110pt}{3pt}{$\hbox to 12pt{\hfill${\scriptstyle 6}$\hfill}$}
\ins{123pt}{3pt}{$\hbox to 12pt{\hfill${\scriptstyle 7}$\hfill}$}
\ins{136pt}{3pt}{$\hbox to 12pt{\hfill${\scriptstyle 8}$\hfill}$}
\ins{149pt}{3pt}{$\hbox to 12pt{\hfill${\scriptstyle 9}$\hfill}$}
\ins{161pt}{3pt}{$\hbox to 12pt{\hfill${\scriptstyle 10}$\hfill}$}
\ins{174pt}{3pt}{$\hbox to 12pt{\hfill${\scriptstyle 11}$\hfill}$}
\ins{187pt}{3pt}{$\hbox to 12pt{\hfill${\scriptstyle 12}$\hfill}$}
\ins{200pt}{3pt}{$\hbox to 12pt{\hfill${\scriptstyle 13}$\hfill}$}
\ins{213pt}{3pt}{$\hbox to 12pt{\hfill${\scriptstyle 14}$\hfill}$}
\ins{225pt}{3pt}{$\hbox to 12pt{\hfill${\scriptstyle 15}$\hfill}$}
\ins{238pt}{3pt}{$\hbox to 12pt{\hfill${\scriptstyle 16}$\hfill}$}
\ins{251pt}{3pt}{$\hbox to 12pt{\hfill${\scriptstyle 17}$\hfill}$}
\ins{264pt}{3pt}{$\hbox to 12pt{\hfill${\scriptstyle 18}$\hfill}$}
\ins{277pt}{3pt}{$\hbox to 12pt{\hfill${\scriptstyle 19}$\hfill}$}
\ins{289pt}{3pt}{$\hbox to 12pt{\hfill${\scriptstyle 20}$\hfill}$}
\ins{0pt}{33pt}{$\hbox to 26pt{\hfill${\scriptstyle -0.200}$\hfill}$}
\ins{0pt}{53pt}{$\hbox to 26pt{\hfill${\scriptstyle -0.150}$\hfill}$}
\ins{0pt}{73pt}{$\hbox to 26pt{\hfill${\scriptstyle -0.100}$\hfill}$}
\ins{0pt}{93pt}{$\hbox to 26pt{\hfill${\scriptstyle -0.050}$\hfill}$}
\ins{0pt}{114pt}{$\hbox to 26pt{\hfill${\scriptstyle -0.000}$\hfill}$}
\ins{0pt}{134pt}{$\hbox to 26pt{\hfill${\scriptstyle 0.050}$\hfill}$}
\ins{0pt}{154pt}{$\hbox to 26pt{\hfill${\scriptstyle 0.100}$\hfill}$}
\ins{0pt}{174pt}{$\hbox to 26pt{\hfill${\scriptstyle 0.150}$\hfill}$}
\ins{0pt}{194pt}{$\hbox to 26pt{\hfill${\scriptstyle 0.200}$\hfill}$}
\ins{308pt}{183pt}{$\hbox to 26pt{\hfill${\scriptstyle 0.000}$\hfill}$}
\ins{308pt}{197pt}{$\hbox to 26pt{\hfill${\scriptstyle 0.002}$\hfill}$}
\ins{312pt}{169pt}{$\hbox to 26pt{\hfill${\scriptstyle -0.002}$\hfill}$}
\ins{312pt}{156pt}{$\hbox to 26pt{\hfill${\scriptstyle -0.004}$\hfill}$}
\ins{312pt}{142pt}{$\hbox to 26pt{\hfill${\scriptstyle -0.006}$\hfill}$}
}{figliap0}{}
\vskip 0.5cm
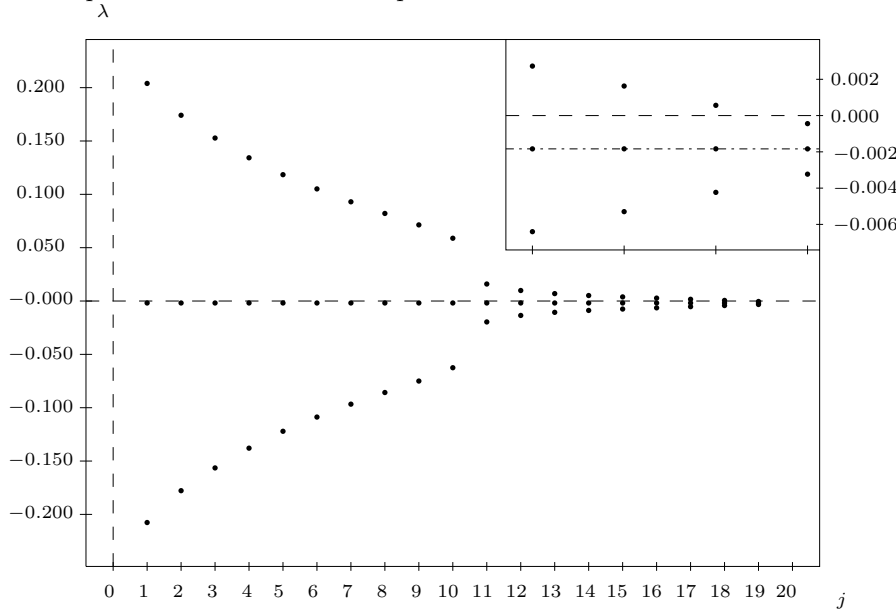
\didascalia{\it Fig.3 The
$38$ Lyapunov exponents, $N=10$. Small picture
magnifies the tail of the larger, showing better
the pairing rule and that the $19$--th exponent is slightly
negative, from \ref{BGG}{}{14}.}
 
\*
In fact a study of the exponents values in the case of $10$ particles
and semiperiodic bounday conditions gives the above diagram in which
the Lyapunov exponents corresponding to $E=1$ are drawn as a function of
the field $E$.  Which shows a very small exponent which is the higher
member of a pair and, yet, is negative.  The graphs of Fig.  2, however,
show that the agreement of \equ(4.1) with the experiment is within the
errors: had there been no negative exponents we would have expected a
slope $1$.  If there is one negative exponent in excess we expect a
slope $1-\fra1{19}$ which is within the error bars in Fig.2. An excess
of $2$ exponents would yield a slope of $1-\fra2{19}$ which is {\it out}
of the error bars.
 
Note that since the exponent smallest in modulus is so small we must
expect that it yields a clear effect only after extremely long times
have elapsed (\ie for values of $\t>4.\,10^3$: totally out of
computability).  At larger fields the number of negative pairs
increases considerably: but the fluctutation theorem cannot be tested
as the fluctuations become too improbabble to be measurable, even with
the largest computers.  \*
 
The reversible case of example 2) has not been studied.  Nor the
reversible case of model 3).  Studies are being performed at least for
the model in example 3).  \*
 
More generally, however, {\it we do not expect the pairing rule to
hold}. For instance this is clear in the case of the second example in
\S2: in that case the Lyapunov exponents relative to the thermostat are
obviously paired to $0$ while those of the conduction particles have a
negative total sum (by the H--theorem of Ruelle, \ref{R}{3}{16}, quoted
before \equ(3.3)).
 
Nevertheless some kind of pairing might still occurr.
In such cases one could envisage that \equ(5.1) is replaced
by a relation like:
 
$$\fra{\l_j^++\l_j^-}2=\media{\s}_+\,\fra{c_j}{2\lis D}\Eq(5.4)$$
where $\media{\s}$ is the $\m_+$ average of the phase space contraction
per unit time; and $c_j$ is some suitable function of $j$, while $\lis
D\DEF \sum_j c_j$.
 
We therefore {\it define} $c_j$ by the \equ(5.4) without attempting at
determining them {\it a priori}. Then one can think that \equ(5.4) holds
in a "almost local" form {\it in the sense that on a rapid time scale
\equ(5.4) becomes true also for the local exponents}.  This means that,
{\it up to an error that tends to zero very quickly with the time $\t$},
the logarithms of the eigenvalues of the matrix
$(J_\t^T(x)J_\t(x))^{1/2\t}$, with $J_\t(x)$ being the jacobian matrix
for the evolution operator $V_\t$ at $x$, divided by $\t$ verify
$\fra12(\l^+_j+\l^-_j)= c_j \b_\t(x)$ with $\b_\t(x)$ denoting the
average $\fra1{2\lis D\t}\sum_{j=-\t/2}^{\t/2-1}\b(S^jx)\,dt$.
 
This property, together with the Axiom C assumption, will then suffice
to extend, in a suitable form, the validity of the predictions of the
fluctuation theorem based on the pairing rule (\ie to cases in which the
attractor is smaller than phase space).
 
We first remark that the really relevant feature of the pairing rule, as
far as the above applications are concerned, is not the constancy of the
pairing {\it but, rather, the fact that some kind of pairing takes place
on a fast enough time scale}, see \ref{Ga}{7}{10}.
 
Then if a local time reversal exists on the attractor (\ie if the
geometric Axiom C is assumed as well for the dynamics the fluctuations
of the observable $\sigma$ will have a "free energy" (or a "generalized
sum of Lyapunov exponents" to adhere to the terminology in
\ref{FP}{}{25}, \ref{BJPV}{}{26}) $\z(p)$, in the sense of \equ(3.4),
with an odd part $p\,\lis{{P}}\, \media{\s}_+$, with $\lis{{P}}$ equal
to $1- \fra{{\sum}_- c_j}{\sum c_j}$ where the $\sum_-$ runs ver the
values of the $j$'s to which correspond two negative Lyapunov exponents:
 
$$\fra{\z(-p)-\z(p)}{\media{\s}_+ p}=\lis P\DEF
1- \fra{{\sum}_- c_j}{\sum c_j}\Eq(5.5)$$
This is a property whose validity can be conceivably tested in, real or
numerical, experiments.  At least the linearity in $p$ of
$-(\z(p)-\z(-p))/\media{\s}_+$ with a slope $\le1$ should be observable.
\*
 
Another application is the already mentioned relation between the
fluctuation theorem and the Onsager reciprocity.
 
We imagine a system even more general than the ones in \S2. We assume
that it is a {\it reversible dissipative} system with several forces
$\V G=(G_1,\ldots,G_s)$ acting upon the particles (or on the fluid in
case of fluid mechanics models).
 
The $\V G$ are parameters, with dimension not necessaribly being that of
a force, measuring the strength of the various causes of nonequilibrium
(\eg they could be electromotive fields as abve, or temperature
differences or other as in the references \ref{Ga}{4}{27}). We follow here
the analysis in \ref{Ga}{5}{28} rather than the one in \ref{Ga}{4}{27}.
 
We suppose that the phase space contraction rate can be written in
increasing powers of $\V G$:
 
$$ \sigma(x)=\sum_{i=1}^s G_i J^0_i(x)+ O(G^2)\Eq(5.6)$$
this simply means that we suppose that at zero forcing, $\V G=\V0$,
there is no phase space contraction and, in fact, we always suppose
that in such case the system is conservative (although this is not
really necessary) and $\m_+|_{\V G=\V0}$ is time reversal invariant.
 
The interpretation of $\s(x)$ as a microscopic density of entropy
production allows us to establish in a unambiguous way the duality
between {\it thermodynamic fluxes} or {\it currents} and {\it
thermodynamic forces}.
 
We define the {\it current} associated with the {\it force} $G_i$ by
$J_i(x)=\dpr_{G_i}\s(x)$ and the {\it transport coefficients} by
setting $L_{ij}=\dpr_{G_j}\langle J_i(x)\rangle_+|_{\V G=0}$ and we
study $L_{ij}$.  We want to show that the above ideas suffice to prove
Onsager's reciprocal relations {\it i.e.} $L_{ij}=L_{ji}$.
 
The analysis is based on the remark that the fluctuation theorem can be
{\it extended} to give properties of the {\it joint} distribution of
the average of $\s$, \equ(5.6), and of the corresponding
$\m_+$--average of $G_j\dpr_{G_j}\s$.
 
In fact we shall define the {\it dimensionless } current associated with
the force $G_j$ by $q=q(x)$:
 
$${1\over\tau} \int_{-\tau/2}^{\tau/2} G_j\dpr_{G_j}
\s(S_tx)dt\,{\buildrel def\over =}\, G_j\langle \dpr_{G_j}\s\rangle_+\,
q\Eq(5.7)$$
where the factor $G_j$ is introduced here only to keep $\s$ and
$G_j\dpr_{G_j}\s$ with the same dimensions.
 
Then if $\pi_\tau(p,q)$ is the joint probability of $p,q$ the {\it
same} proof of the fluctuation theorem discussed above yields also
that, if we define:
 
$$
\lim_{\tau\to\infty}
{1\over\tau}\log\pi_\tau(p,q)=-\zeta(p,q)\Eq(5.8)$$
then:
$$\fra{\z(-p,-q)-\z(p,q)}{p\media{\s}_+}=1\Eq(5.9)$$
The proof in \ref{Ga}{2}{17} shows that $\z(p,q)$ is analytic in the
interior of the domain of variability of $p,q$ at least if this domain
is open.  The latter condition means that $p,q$ are {\it independent
variables} for all $\V G$ small enough.
 
This is the "normal" case: it is violated essentially only if the forces
are not really different although they are given different names, \eg
one could have only one force $G_1$ acting on the system but one could
regard it as two equal forces.  In this case $\s$ and $G_1\dpr_{G_1}\s$
would coincide (at least to first order in $\V G$) and the $p,q$ would
not be independent.
 
We can compute $\zeta(p,q)$ as usual in statistical mechanics:
by considering first the transform $\lambda(\beta_1,\beta_2)$:
 
$$\l(\V\b)=
\lim_{\tau\to\infty}
{1\over\tau}\log \int e^{\tau(\beta_1\, (p-1)
\langle\s\rangle_++
\beta_2\,(q-1)\langle G_j\dpr_{G_j}\s\rangle_+)}\pi_\tau(p,q)dpdq
\Eq(5.10)$$
and then the Legendre trasform:
 
$$\zeta(p,q)=\max_{\beta_1,\beta_2}\big(
\beta_1\, (p-1)\langle\s\rangle_++
\beta_2\,(q-1)
\langle G_j\dpr_{G_j}\s\rangle_+-\l(\V\b)\big)\Eq(5.11)$$
The function $\lambda(\V\beta)$, $\V\b=(\b_1,\beta_2)$, is evaluated by
the {\it cumulant expansion}.  For the purpose of shortening the
notation introduce $X(x)$ as:
 
$$X(x)=\beta_1\sum_{t=-\tau/2}^{\tau/2-1}(\s(S^tx)
-\media{\s}_+)\,+\beta_2\,\sum_{t=-\tau/2}^{\tau/2-1}(G_j\dpr_j\s(S^tx)
-\langle G_j\dpr_j\s\rangle_+)\Eq(5.12)$$
so that $X$ has $\mu_+$--average $0$ and the function $\lambda(\V\beta)$
is simply written as:
$$e^{\tau \lambda(\V\beta)}
=\int e^{X(x)} \mu_+(dx)=\langle{ e^X}\rangle_+\Eq(5.13)$$
But $\langle X\rangle_+=0$ by the definition so that the cumulant
expansion for $\int e^{X(x)} \mu_+(dx)$ yields, forgetting $O(G^3)$:
$$\int e^{X(x)} \mu_+(dx)=e^{{1\over{2!}}\langle X^2\rangle_+}\Eq(5.14)$$
and $\langle X^2\rangle_+$ is a quadratic form $C\V\b,\V\b)$ in $\V\b$
(because $X$ is linear in $\V\beta$) with coefficients given by the
cumulants:
 
$$\eqalign{
C_{11}=&\sum_{-\io}^\io \big(
\media{\s(S^t\cdot)\s(\cdot)}_+-
\media{\s(\cdot)}_+^2\big)\cr
C_{22}=&\sum_{-\io}^\io \big(
\media{G_j\dpr_{G_j}\s(S^t)\cdot G_j\dpr_{G_j}\s(\cdot)}_+-
\media{G_j\dpr_{G_j}\s(\cdot)}_+^2\big)\cr
C_{12}=&C_{21}=\sum_{-\io}^\io \big(
\media{\s(S^t)\cdot G_j\dpr_{G_j}\s(\cdot)}_+-
\media{\s}_+\,\media{G_j\dpr_{G_j}\s(\cdot)}_+\big)\cr}\Eq(5.15)$$
The four cumulants form a symmetric matrix that will be called $C$.
 
The functions $\l(\V\b)$ and $\z(p,q)$ are related via a Legendre
transform:
$$\z(p,q)=\max_{\V\b}\big(\V\b\cdot\V w-\l(\V\b)\big)\Eq(5.16)$$
where $\V w=\pmatrix{
(p-1)\langle \s\rangle_+\cr
(q-1)\langle G_j\dpr_j\s\rangle_+\cr}$.
 
So that the maximum condition yields the equations for the value
of $\V\b$ where the expression in parenthesis in \equ(5.16) reaches the
maximum $\V\b=\V \beta_{\max}$:
 
$$\eqalign{ (p-1)\langle \s\rangle_+=&\dpr_{\beta_1}
\lambda(\V\beta)|_{\V\beta=\V\beta_{\max}}\cr (q-1)\langle
G_j\dpr_j\s\rangle_+=& \dpr_{\beta_2}
\lambda(\V\beta)|_{\V\beta=\V\beta_{\max}}\cr}\Eq(5.17)$$
 
so that, always performing the computations by neglecting quantities of
$O(G^3)$ (\ie by using $\l(\V\b)=\fra12(C\V\b,\V\b)$):
 
$$\V w= \pmatrix{ (p-1)\langle \s\rangle_+\cr (q-1)\langle
G_j\dpr_j\s\rangle_+\cr}= \pmatrix{C_{11}& C_{12}\cr C_{21}& C_{22}\cr}
\pmatrix{\beta_1\cr\beta_2\cr}\qquad {\rm or}\quad
\V\beta_{\max}=C^{-1}\V w\Eq(5.18)$$
and from \equ(5.16), evaluating the r.h.s. at the maximum point
$\V\beta_{\max}$, we get:
 
$$\zeta(p,q)=\V w\cdot\V\beta_{\max}-\lambda(\V\beta_{\max})=
{1\over2}(\V w,C^{-1}\V w)\Eq(5.19)$$
 
Since:
 
$$C^{-1}={1\over\det C}\pmatrix{C_{22} &-C_{12}\cr
-C_{21}& C_{11}\cr}\Eq(5.20)$$
we see that (recall that $C$ is symmetric):
$$\eqalign{ &\zeta(p,q)={1\over2}
(C^{-1})_{11}(p-1)^2\langle\s\rangle_+^2+\cr
&+{1\over2}(C^{-1})_{22}(q-1)^2\langle G_j\dpr_js\rangle^2_+
+(C^{-1})_{12}(q-1)(p-1)\s_+\,G_j\dpr_j\s_+\cr}\Eq(5.21)$$
and the odd terms in $(p,q)$ have coefficients:
$$\eqalign{
q\quad\to\quad
&(-2(C^{-1})_{22}G_j\media{\dpr_{G_j}\s}_+
-2(C^{-1})_{12}\media{\s}_+)\,G_j\media{\dpr_{G_j}\s}_+=0\cr
p\quad\to\quad
&(-2(C^{-1})_{11}\media{\s}_+
-2(C^{-1})_{12}G_j\media{\dpr_{G_j}\s}_+)
\media{\s}_+)=-\media{\s}_+\cr}\Eq(5.22)$$
where the r.h.s. {\it arise by imposing compatibility with the
fluctuation theorem relation} \equ(5.9).
 
The above two relations, after some simple algebra, are seen to imply:
 
$$\eqalign{ \langle\s\rangle_+=&{1\over 2}C_{11}+ O(G^3)\cr \langle
G_j\dpr_{G_j}\s\rangle_+=&{1\over2}C_{12}+O(G^3)\cr}\Eq(5.23)$$
Then expanding both sides of \equ(5.23) {\it to lowest order} (\ie
lowest non trivial, the second in this case) in the $G_i$'s we get the
Onsager relation and the Green Kobo formula.
 
In fact first we look at the r.h.s.  of the first relation in
\equ(5.23): in $C_{11}$ one simply replaces $\s$, see \equ(5.15),
by its expansion to first order, \equ(5.6), and the r.h.s. becomes a
quadratic form in $\V G$ with coefficients given by:
 
$${1\over 2}\int_{-\infty}^\infty dt\, \big(\langle J^0_i(S_t\cdot)
J^0_j(\cdot)\rangle_+- \langle J^0_i\rangle_+\langle J^0_j\rangle_+
\big)\big|_{G=0}\Eq(5.24)$$
On the other hand the expansion of $\media{\s}_+$ in the l.h.s. of
the first of \equ(5.23) to second order in $\V G$ gives:
 
$$\media{\s}_+=\sum G_i\dpr_{G_i}\media{\s}_{+}\big|_0+ {1\over2}
\sum_{ij} G_i G_j \big(\dpr_{G_i}(\dpr_{G_j}\media{\s}_+)\big|_0\Eq(5.25)$$
the first term vanishes (by time reversal, or just because it is clear
from the r.h.s.  of \equ(5.23) that $\media{\s}_+$ is of second order in
$\V G$).  The second term is the sum of ${1\over2}G_iG_j$ times:
 
$$\eqalign{ &\dpr_{G_i}\dpr_{G_j} \int \s(x) \mu_+(dx) =\int
\big(\dpr_{G_i}
\dpr_{G_j}\s(x) \mu_+(dx)\,+\cr &+
\,\dpr_{G_i}\s(x)\dpr_{G_j}\mu_+(dx)+\dpr_{G_j}\s(x)\dpr_{G_i}\mu_+(dx)
+\s(x)\dpr_{G_i}\dpr_{G_j}\mu_+(dx)\big)\cr}\Eq(5.26)$$
evaluated at $G=0$; and the first term in the r.h.s.  vanishes (by time
reversal it changes sign, while $\m_+$ is invariant) the second and
third terms are $\dpr_j<{J^0_i}>_+|_0+(i \otto j)$ and the fourth
vanishes (because $\s=0$ at $G=0$): note that
$(\dpr_{G_j}<{J^0_i}>_+)|_0$ is equal to $(\dpr_{G_j}<{J_i}>_+)|_0$
).\annota{3}{\nota This is again the same argument:
 
$$\eqalign{ &\dpr_j<J_i>_+|_0\equiv\dpr_j(\int
\dpr_i\s(x)\mu_+(dx))|_0= (\int \dpr_j
\dpr_i\s(x)\mu_+(dx))_0+ (\int
\dpr_i\s(x)\dpr_j\mu_+(dx))_0=\cr &=0+ (\int
\dpr_i\s(x)|_0 \dpr_j\mu_+(dx)|_0=
\dpr_j<J^0_i>_+)_0\cr}$$
where the last step uses that $J^0_i$ being evaluated at $G=0$ does not
depend on $G$.}
 
Therefore by equating the r.h.s and the l.h.s.of \equ(5.23), we get from
the first of \equ(5.23) the above expression \equ(5.24) for the matrix
$\fra{L_{ij}+L_{ji}}2$ giving Green--Kubo's formula for $i=j$ (but not
the Onsager reciprocity nor the general Green--Kubo formula which would
say that $L_{ij}$ equals \equ(5.24)).
 
The same type of analysis on the second of \equ(5.23), {\it which,
unlike the first relation, is not symmetric in the $\V G$'s as $j$ is
privileged}, leads to the "asymmetric relation'':
 
$$L_{ji}\=\dpr_{G_j}\media{\dpr_{G_i}\s}|_{\V G=\V0}= {1\over
2}\sum_{t=-\io}^\io \big(\media{J^0_i(S_t\cdot) J^0_j(\cdot)}_+-
\media{J^0_i}_+\media{J^0_j}_+ \big)\big|_{G=0}\Eq(5.27)$$
which gives the general Green-Kubo formula, hence Onsager reciprocity
and the fluctuation dissipation theorem.
 
Thus the Onsager relations are a consequence of the fluctuation theorem
(not surprisingly) and of its (obvious) extension, \equ(5.9), in the
limit $G\to0$, when combined with the cumulant expansion for entropy
fluctuations.  Those theorems and the fast decay of the $\s\-\s$
correlations are all natural consequences of the chatic hypothesis in
reversible statistical mechanical or fluid mechanical systems.
 
Therefore while the Onsager reciprocity and Green--Kubo formulae (or
fluctuation dissipation theorem) only hold around equilibrium, {\it
i.e.} they are properties of $G$--derivatives evaluated at $G=0$, and
the cumulant expansion for $\lambda(\V\beta)$ is a general consequence
of the correlation decay in Anosov systems, the fluctuation theorem also
holds far from equilibrium, {\it i.e.} for large $\V G$, and can be
considered a generalization of the Onsager relations and of the
Green--Kubo formulae, \ref{Ga}{5}{28}.
 
Evidence for the above interpretation of the fluctuation theorem arose
in \ref{BGG}{}{14} in an effort to interpret the results of various
numerical experiments and an apparent incompatibility of the {\it a
priori} known non gaussian nature of the distribution $\pi_\tau(p)$ and
the "gaussian looking" empirical distributions.
 
\*
The next question is what we can say about the irreversible models?
The key is the already mentioned conjectured equivalence of the {\it
corresponding} SRB distributions.
 
The equivalence conjectures seem to raise the possibility of a genral
theory of non equilibrium statistical ensembles and of their
equivalence.
 
The idea of using reversible equations to study rreversible behaviour is
quite appealing as we have seen that there are by now quite a few
results that can be obtained for reversible dissipative evolutions.
Hopefully the theory will grow and we shall learn to use the chaotic
hypothesis and the ensuing characterization of the SRB distributions.
After all in equilibrium statistical mechanics it took a long time to
develop the theory from the heat theorem to the universal phenomena at
criticality.
 
Unfortunately the equivalence of the ensembles does not allow us to make
predictions based on the fluctuation theorem about the fluctuations of
the entropy production in irreversible systems, even when equivalent to
reversible ones. This is so because the entropy production is a "global"
observable that in the irreversible models is ususally fixed \ap to a
given value or, at least, to a value related to a quantity fixed {\it a
priori}.
 
Thus the fact that we cannot translate knowledge of the entropy
fluctuations in reversible models to irreversible equivalent models
becomes analogous to the impossibility of translating information on the
energy fluctuations in the canonical enseble to informations (of any
interest) on the energy fluctuations in the microcanonical ensemble.
 
Thus although equivalence can be tested, we have, so far, no special
predictions to mention aside the obvious one that one should get the
same result when measuring corresponding quantities (a rather non
trivial and interesting property): the fluctuation theorem is not
directly usable beyond the fact that it implies the Onsgaer reciprocity.
 
In \ref{Ga}{7}{10} some applications have been proposed for the analysis of
the models in the example 3) of \S2 and they may be tested, perhaps,
experimentally. But this is beyond the scopes of the present review.
 
The equivalence conjecture {\it on the other hand} gives us an
opportunity to test how good the approximations of irreversible
equations for macroscopic phenomena are.
 
In fact if a measurement (a hypothtical one as so far no measurements of
this type seem to have been performed) of the entropy fluctuations
agrees with the fluctuation theorem then this would mean that the
reversible models are better models (compared to the irreversible ones)
of the system as they give the same predictions for normal observables
but for the entropy production fluctuations they give different
predictions (compared to the ones of the irreversible models).
 
\*
\0{\it Appendix A1.
An example of Axiom C system.} \numsec=1\numfor=1\*
 
We give here an example, taken from \ref{BG}{}{6}, in which $i^*$, the
local time reversal, arising in the applications can be easily
constructed.  The example illustrates what we think is a typical
situation.  The poles $\O_\pm$, $\O_-$, will be two compact regular
surfaces, identical in the sense that they will be mapped into each
other by the time reversal $i$ defined below.
 
If $x$ is a point in $M_*=\O_+$ the generic point of
the phase space will be determined by a pair $(x,z)$ where $x\in M_*$
and $z$ is a set of transversal coordinates that tell us how far we are
from the attractor.  The coordinate $z$ takes two well defined values
on $\O_+$ and $\O_-$ that we can denote $z_+$ and $z_-$ respectively.
 
The coordinate $x$ identifies a point on the compact manifold $M_*$ on
which a reversible transitive Anosov map $S_*$ acts (see \ref{Ga}{3}{18}).
And the map $S$ on phase space is defined by:
 
$$S(x,z)=(S_*x,\tilde S z)\Eqa(A1.1)$$
where $\tilde S$ is a map acting on the $z$ coordinate (marking a point
on a compact manifold) which is an evolution leading from an unstable
fixed point $z_-$ to a stable fixed point $z_+$.  For instance $z$
could consist of a pair of coordinates $v,w$ with $v^2+w^2=1$ (\ie $z$
is a point on a circle) and an evolution of $v,w$ could be governed by
the equation $\dot v=-\a v, \ \dot w= E-\a w$ with $\a={Ew}$.  If we
set $\tilde Sz$ to be the time $1$ evolution (under the latter
differential equations) of $z=(v,w)$ we see that such evolution sends
$v\to0$ and $w\to\pm 1$ as $t\to\pm\io$ and the latter are non marginal
fixed points for $\tilde S$.
 
Thus if we set $S(x,z)=(S_*x,\tilde S z)$ we see that our system is
hyperbolic on the basic sets $\O_\pm=M_*\times \{z_\pm\}$ and the
future pole $\O_+$ is the set of points
$(x,z_+)$ with $x\in M_*$; while the past pole $\O_-$
is the set of points $(x,z_-)$ with $x\in M_*$.
 
Clearly the two poles are mapped into each other by the map
$i(x,z_{\pm})=(i^*x,z_{\mp})$.  But on each attractor a "local time
reversal" acts: namely the map $i^* (x,z_\pm)=(i^*x,z_\pm)$.
 
The system is "chaotic" as it has an Axiom A attracting set with closure
consisting of the points having the form $(x, z_+)$ for the motion
towards the future and a different Axiom A attracting set with closure
consisting of the points having the form $(x, z_-)$ for the motion
towards the past. In fact the dynamical systems $(\O_+,S)$ and
$(\O_-,S)$ obtained by restricting $S$ to the future or past
attracting sets are Anosov systems because $\O_\pm$ are regular
manifolds.
 
We may think that in the reversible cases the situation is always the
above: namely there is an "irrelevant" set of coordinates $z$ that
describes the departure from the future and past attractors. The future
and past attractors are copies (via the global time reversal $i$) of
each other and on each of them is defined a map $i^*$ which inverts the
time arrow, {\it leaving the attractor invariant}: such map will be
naturally called the {\it local time reversal}.
 
In the above case the map $i^*$ and the coordinates $(x,z)$ are
"obvious".  The problem is to see that they are defined quite generally
under the only assumption that the system is reversible and has a
unique future and a unique past attractos that verify the Axiom A.
This is a problem that is naturally solved in general when the system
verifies the Axiom C of \S1.
 
In the following section we shall describe the interpretation of $i^*$
in terms of symbolic dynamics when the system verifies Axiom C: as one
may expect the construction is simple but it is deeply related to the
properties of hyperbolic systems such as their Markov partitions.
\*
 
\0{\it Appendix A2. Ising model analogy.} \numsec=2\numfor=1\*
 
The fluctuation theorem as expressed by \equ(2.10) and the subsequent
comments on the Gaussian nature of the function $\p_\t(p)$ may seem
somewhat strange and unfamiliar.
 
It is therefore worth pointing out that the phenomenon of a ``linear
fluctuation law'' on the odd part of the distribution, in the sense of
\equ(4.1), without a globally Gaussian distribution, is in fact well
known in statistical mechanics and probability theory. And an example,
proposed in \ref{BGG}{}{14}, of what the fluctuation theorem means in a
concrete case, in which $\p_\t(p)$ is {\it not} Gaussian, can be made
by using the Ising model on a $1$ dimensional lattice $Z$.
 
We consider the space $\CC$ of the {\it spin configurations} $\V
\s=\{\s_\x\}, \,\x\in Z$ and the map $S$ that translates each
configuration to the right (say). The ``time reversal'' is the map
$i:\{\V\s\}\to\{-\V \s\}$ that changes the sign to each spin.
 
The probability distribution that approximates the SRB distribution is
the finite volume Gibbs distribution:
 
$$\m_\L(\V \s)=\fra{\exp\big(J\sum_{j=-T}^{T}\s_j\s_{j+1}\,+\,
h\sum_{j=-T}^T \s_j\big)}{normalization} \Eqa(A2.1)$$
where $\L=[-T,T]$ is a large interval, $J, h>0$. The configuration $\V
\s$ outside $\L$ is distributed independently on the one inside the box
$\L$, to fix the ideas.
 
Calling $\media{m}_+$ the average magnetization in the thermodynamic
limit we define the magnetization in a box $[-\fra\t2,\fra\t2]$ to be
$M_\t=\t \media{m}_+ p$ and we look at the probability distribution
$\p_\t^T(p)$ of $p$ in the limit $T\to\io$. The Gibbs distribution
corresponding to the limit of \equ(A2.1) will play the role of the SRB
distribution. Calling this limit probability $\p_\t(p)$ it is easy to
see that:
 
$$\fra{\p_\t(p)}{\p_\t(-p)}\,\tende{\t\to\io}\,e^{2\t h\media{m}_+\,p}
\Eqa(A2.2)$$
This is in fact obvious if we take the two limits $T\to\io$ and
$\t\to\io$ {\it simultaneously} by setting $T=\fra\t2$. In such a case,
if $\sum_{\V\s;\,p}$ denotes summation over all the configurations with
given magnetization in $[-T,T]$, \ie such that
$\sum_{j=-\fra\t2}^{\fra\t2-1}\s_j=\media{m}_+ \,p$
the distribution \equ(A2.1) gives us immediately that:
 
$$
\fra{\p^T_\t(p)}{\p^T_\t(-p)}=
\fra{\sum_{\V\s,\, p}\exp{J\sum_{j=-T}^{T}\s_j\s_{j+1}\,+\,
h\sum_{j=-T}^T \s_j}}{\sum_{\V\s,\, -p}\exp{J
\sum_{j=-T}^{T}\s_j\s_{j+1}\,+\,h\sum_{j=-T}^T \s_j}}
\=e^{2\t h\media{m}_+\,p}\Eqa(A2.3)$$
if we use the symmetry of the pair interaction part of the energy under
the ``time reversal'' (\ie under spin reversal).
 
The error involved, in the above argument, in taking $T=\fra\t2$ rather
then first $T\to\io$ and then $\t\to\io$, can be easily corrected since
the corrections are ``boundary terms'', and in one dimensional short
range spin systems there are no phase transitions and the boundary terms
have no influence in the infinite volume limit (\ie they manifest
themselves as corrections that vanish, as $T\to\io$ followed by
$\t\to\io$).
 
One may not like that the operation $i$ commutes with $S$ rather than
transforming it into $S^{-1}$. Another example in which the operation
$i$ does also invert the sign of time is obtained by defining $i$ as
$\{i\V\s\}|_j= -\s_{-j}$: the \equ(A2.3) can be derived also by using this
new symmetry operation.
 
The above examples show why there is \ap independence between any
Gaussian property of $\p_\t(p)$ and the fluctuation theorem. The theory
of the fluctuation theorem in \ref{GC}{1}{13} is in fact {\it based} on
the possibility (discovered in \ref{S}{}{2}) of representing a chaotic
system as a one dimensional short range system of interacting spins (in
general higher that $\fra12$); and the argument is, actually, very close
to the above one for the Ising model with, however, a rather different
time reversal operation. See \ref{Ga}{2}{17} for mathematical details on
the boundary condition question.
\*

\0{\it Acknowledgements:} I are indebted to F.  Bonetto, E.G.D.  Cohen,
P.  Garrido, G .  Gentile, G.  Paladin for stimulating discussions and
comments: the results reviewed here can be originally found in joint
works or originated from common discussions. This work is an expanded
version of a series of talks at the meeting {\it Let's face chaos
through nonlinear dynamics}, at the University of Maribor, Slovenia, 24
june 1996-- 5 july 1996.  It is part of the research program of the
European Network on: "Stability and Universality in Classical
Mechanics", \# ERBCHRXCT940460, and it has been partially supported
also by CNR-GNFM and Rutgers University.
 
\vskip1truecm%
 
\0{\bf References.}
\vskip1truecm
\def\aAA{ Arnold, V., Avez, A.: {\it Ergodic problems of classical
mechanics}, Benjamin, 1966.}
\rif{AA}{}{\aAA}{0}
 
\def\aBG{ Bonetto, F., Gallavotti, G.: {\it Reversibility,
coarse graining and the chaoticity principle},  mp$\_$arc@ math. utexas.
edu, \#96--50.}
\rif{BG}{}{\aBG}{0}
 
\def\aBGG{ Bonetto, F., Gallavotti, G., Garrido, P.: {\it Chaotic
principle: an experimental test}, mp$\_$arc @math. utexas. edu,
\# 96-154.}
\rif{BGG}{}{\aBGG}{}
 
\def\aBJPV{Bohr, T., Jensen M.H., Paladin, G., Vulpiani, A.:
{\it Dynamical systems approach to turbulence}, Cambridge Nonlinear
series, 1996.}
\rif{BJPV}{}{\aBJPV}{0}

\def\aCELS{ Chernov, N. I., Eyink, G. L., Lebowitz, J.L., Sinai, Y.:
{\it Steady state electric conductivity in the periodic Lorentz gas},
Communications in Mathematical Physics, {\bf 154}, 569--601, 1993.}
\rif{CELS}{}{\aCELS}{0}
 
\def\aDM{ Dettman, C.P., Morriss, G.P.: {\it Proof of conjugate pairing
for an isokinetic thermostat}, N.S. Wales U., Sydney 2052, preprint,
1995.}
\rif{DM}{}{\aDM}{0}
 
\def\aDGM{ de Groot, S., Mazur, P.: {\it Non equilibrium thermodynamics},
Dover, 1984.}
\rif{DGM}{}{\aDGM}{0}
 
\def\aDPH{ Dellago, C., Posch, H., Hoover, W.: {\it Lyapunov instability in
system of hard disks in equilibrium and non-equilibrium steady states},
Physical Review, {\bf 53E}, 1485--1501, 1996.}
\rif{DPH}{}{\aDPH}{0}
 
\def\aEM{Evans, D.J., Morriss, G.P.: {\it Statistical Mechanics of
Nonequilibrium fluids}, Academic Press, New York, 1990.}
\rif{EM}{}{\aEM}{0}

\def\aECM{ Evans, D.J.,Cohen, E.G.D., Morriss, G.P.: {\it Viscosity of a
simple fluid from its maximal Lyapunov exponents}, Physical Review, {\bf
42A}, 5990--\-5997, 1990.}
\rif{ECM}{1}{\aECM}{0}
 
\def\bECM{ Evans, D.J.,Cohen, E.G.D., Morriss, G.P.: {\it Probability
of second law violations in shearing steady flows}, Physical Review
Letters, {\bf 71}, 2401--2404, 1993.}
\rif{ECM}{2}{\bECM}{0}

\def\aFP{Frisch, U., Parisi, G.: in {\sl Turbulence and predictability},
p. 84--87, ed. M. Ghil, R. Benzi, G.Parisi, Noth Holland, 1984.}
\rif{FP}{}{\aFP}{0}

\def\aGa{ Gallavotti, G.: {\it Ergodicity, ensembles, irreversibility
in Boltzmann and beyond}, Journal of Statistical Physics, {\bf 78},
1571--1589, 1995.}
\rif{Ga}{1}{\aGa}{0}
 
\def\bGa{ Gallavotti, G.: {\it
Reversible Anosov diffeomorphisms and large deviations.},
Ma\-the\-ma\-ti\-cal Physics Electronic Journal, {\bf 1}, (1), 1995,
(http:// www. ma. utexas. edu/ MPEJ/ mpej.htlm).}
\rif{Ga}{2}{\bGa}{0}
 
\def\cGa{ Gallavotti, G.: {\it Topics in chaotic dynamics}, Lectures
at the Granada school 1994, Ed. P. Garrido, J. Marro,
in Lecture Notes in Physics, Springer Verlag, {\bf 448}, p. 271--311,
1995.}
\rif{Ga}{3}{\cGa}{0}
 
\def\dGa{ Gallavotti, G.:{\it Chaotic hypothesis: Onsager reciprocity
and fluctuation--dis\-si\-pa\-tion theorem}, J. Statistical Physics,
{\bf 84}, 899--926, 1996. In print in Journal of
Statistical Physics, 1996. See also: {\it Chaotic principle: some
applications
to developed turbulence}, archived in {\it mp$\_$arc@math.utexas.edu},
\#95-232, and {\it chao-dyn@ xyz.lanl.gov}, \#9505013, in print in J. of
Statistical Physics.}
\rif{Ga}{4}{\dGa}{0}
 
\def\eGa{ Gallavotti, G.: {\it Extension of Onsager's reciprocity to
large fields and the chao\-tic hypothesis},
mp$\_$ arc 96-109; or chao-dyn 9603003, in print in Physical Review Letters.}
\rif{Ga}{5}{\eGa}{0}
 
\def\fGa{ Gallavotti, G.: {\it Equivalence of dynamical ensembles
and Navier Stokes equations}, mp$\_$ arc 96-131; or chao-dyn
9604006, in print on Physics Letters A.}
\rif{Ga}{6}{\fGa}{0}
 
\def\gGa{ Gallavotti, G.: {\it
Dynamical ensembles equivalence in statistical mechanics}, in
mp$\_$arc@ math. utexas. edu \#96-182,
chao-dyn@xyz.lanl.gov \#9605006, in print on Physica D.}
\rif{Ga}{7}{\gGa}{}
 
\def\aGC{ Gallavotti, G., Cohen, E.G.D.: {\it Dynamical ensembles in
nonequilibrium statistical mechanics}, Physical Review Letters,
{\bf74}, 2694--2697, 1995.}
\rif{GC}{}{\aGC}{}
 
\def\bGC{ Gallavotti, G., Cohen, E.G.D.: {\it Dynamical ensembles in
stationary states}, Journal of Statistical Physics, {\bf 78}, ,1995.}
\rif{GC}{}{\bGC}{}
 
\def\aHHP{ Holian, B.L., Hoover, W.G., Posch. H.A.: {\it Resolution of
Loschmidt's paradox: the origin of irreversible behavior in reversible
atomistic dynamics}, Physical Review Letters, {\bf 59}, 10--13, 1987.}
\rif{HHP}{}{\aHHP}{0}
 
\def\aLL{ Landau, L., Lifshitz, V.: {\it M\'ecanique des fluides}, MIR,
Moscow, 1966.}
\rif{LL}{}{\aLL}{0}
 
\def\aN{ Nelson, E.: {\sl Dynamical theories of brownian motion},
Mathematical Notes, Princeton U. Press, 1967.}
\rif{N}{}{\aN}{}
 
\def\aR{ Ruelle, D.: {\it Chaotic motions and strange attractors},
Lezioni Lincee, notes by S.  Isola, Accademia Nazionale dei Lincei,
Cambridge University Press, 1989; see also: Ruelle, D.: {\it Measures
describing a turbulent flow}, Annals of the New York Academy of
Sciences, {\bf 357}, 1--9, 1980.  For more technical expositions see
Ruelle, D.: {\it Ergodic theory of differentiable dynamical systems},
Publications Math\'emathiques de l' IHES, {\bf 50}, 275--306, 1980.}
\rif{R}{1}{\aR}{}
 
\def\bR{ Ruelle, D.: {\it A measure associated with Axiom A
attractors}, American Journal of Mathematics, {\bf98}, 619--654, 1976.}
\rif{R}{2}{\bR}{}
 
\def\cR{ Ruelle, D.: {\it Positivity of entropy production in the
presence of a random thermostat}, in mp$\_$arc@ math. utexas. edu,
\# 96-167. See also {\it Positivity of entropy production in
nonequilibrium statistical mechanics}, in mp$\_$arc@ math. utexas. edu,
\# 96-166.}
\rif{R}{3}{\cR}{}
 
\def\dR{ Ruelle, D.: {\sl Elements of differentiable dynamics and
bifurcation theory}, Academic Press, 1989.}
\rif{R}{4}{\dR}{}
 
\def\aS{ Sinai, Y.G.: {\sl Gibbs measures in ergodic theory},
Russian Mathematical Surveys, {\bf 27}, 21--69, 1972.  Also: {\it
Introduction to ergodic theory}, Prin\-ce\-ton U.  Press, Princeton,
1977.}
\rif{S}{}{\aS}{}
 
\def\aSm{ Smale, S.: {\it Differentiable dynamical systems}, Bullettin
of the American Mathematical Society, {\bf 73 }, 747--818, 1967.}
\rif{Sm}{}{\aSm}{}

\raf{Ga}{1}{\aGa}{1}
\raf{S}{}{\aS}{2}
\raf{Sm}{}{\aSm}{3}
\raf{AA}{}{\aAA}{4}
\raf{R}{4}{\dR}{5}
\raf{BG}{}{\aBG}{6}
\raf{GC}{2}{\bGC}{7}
\raf{Ga}{6}{\fGa}{8}
\raf{N}{}{\aN}{9}
\raf{Ga}{7}{\gGa}{10}
\raf{LL}{}{\aLL}{11}
\raf{R}{2}{\bR}{12}
\raf{GC}{1}{\aGC}{13}
\raf{BGG}{}{\aBGG}{14}
\raf{DGM}{}{\aDGM}{15}
\raf{R}{3}{\cR}{16}
\raf{Ga}{2}{\bGa}{17}
\raf{Ga}{3}{\cGa}{18}
\raf{ECM}{2}{\bECM}{19}
\raf{CELS}{}{\aCELS}{20}
\raf{HHP}{}{\aHHP}{21}
\raf{EM}{}{\aEM}{22}
\raf{ECM}{1}{\aECM}{23}
\raf{DM}{}{\aDM}{24}
\raf{FP}{}{\aFP}{25}
\raf{BJPV}{}{\aBJPV}{26}
\raf{Ga}{4}{\dGa}{27}
\raf{Ga}{5}{\eGa}{28}
 
\*
\0{\it Internet access:
All the Author's quoted preprints can be found and freely downloaded
(latest postscript version including corrections of misprints and
errors) at:
 
\centerline{\tt http://chimera.roma1.infn.it}
 
\0in the Mathematical Physics Preprints page.\\
\sl e-mail address of author: giovanni@ipparco.roma1.infn.it
}
 
\ciao